\documentclass[authoryear,review]{elsarticle}
\makeatletter
\def\ps@pprintTitle{%
 \let\@oddhead\@empty
 \let\@evenhead\@empty
 \def\@oddfoot{}%
 \let\@evenfoot\@oddfoot}
\makeatother

\usepackage[utf8]{inputenc}
\usepackage{fontenc}[t1]

\usepackage{amsmath,amssymb,amsfonts}
\usepackage{mathtools}
\usepackage{fullpage}

\usepackage{fancyhdr}
\usepackage[onehalfspacing]{setspace}
\usepackage{tikz}
\usetikzlibrary{positioning}

\usetikzlibrary{decorations.pathreplacing}
\usepackage{caption, subcaption}

\usepackage{natbib}
\usepackage{graphicx}
\usepackage{epstopdf}
\newtheorem{prop}{Proposition}[section]

\def\x{\tilde{x}}
\def\y{\tilde{y}}
\def\z{\tilde{z}}
\def\vt{\tilde{v}}
\def\d{\partial}
\def\L{L^{(0)}}
\def\LL{L^{(1)}}
\def\r{\rho^{(0)}}
\newcommand{\kommentar}[1]{}

\newcommand{\field}[1]{\mathbb{#1}}
\newcommand {\R}{\field{R} }
\newcommand {\N}{ {\field{N}} }

\newcommand {\eps}{\varepsilon }

\begin{document}

\begin{frontmatter}

\title{Fisher-Wright model with deterministic seed bank and selection}

\author[TUMGarching]{Lukas Heinrich} 
\author[TUMGarching,ICB]{Johannes M\"uller\corref{mycorrespondingauthor}} 
\author[TUMWeihen]{Aur\'elien Tellier} 
\author[TUMWeihen]{Daniel \v Zivkovi\'c}

\address[TUMGarching]{Center for Mathematics, Technische Universit\"at M\"unchen, 85748 Garching, Germany}
\address[TUMWeihen]{Section of Population Genetics, Center of Life and Food Sciences Weihenstephan, Technische Universit\"at M\"unchen, 85354 Freising, Germany}
\address[ICB]{Institute for Computational Biology, Helmholtz Center Munich, 85764 Neuherberg, Germany}

\cortext[mycorrespondingauthor]{Corresponding author}

\title{Effects of population- and seed bank noise on neutral evolution and efficacy of natural selection}


\begin{abstract}
Population genetics models typically consider a fixed population size and 
a unique selection coefficient. However, population dynamics inherently 
generate noise in numbers of individuals and selection acts on various 
components of the individuals' fitness. In plant species with seed banks, 
the size of both the above- and below-ground compartments present noise 
depending on seed production and the state of the seed bank. We investigate 
if this noise has consequences on 1)~the rate of genetic drift, 
and 2)~the efficacy of selection. We consider four variants of two-allele 
Moran-type models defined by combinations of presence and absence of noise 
in above-ground and seed bank compartments. Time scale analysis and dimension 
reduction methods allow us to reduce the corresponding Fokker-Planck equation 
to a one-dimensional diffusion approximation of a Moran model. 
We first show that if 
the above-ground noise classically affects the rate of genetic drift, 
below-ground noise reduces the diversity storage effect of the seed bank. 
Second, we consider that selection can act on four different components 
of the plant fitness: plant or seed death rate, seed production or seed 
germination. Our striking result is that the efficacy of selection for seed 
death rate or germination rate is reduced by seed bank noise, whereas 
selection occurring on plant death rate or seed production is not affected. 
We derive the expected site-frequency spectrum reflecting this heterogeneity 
in selection efficacy between genes underpinning different plant fitness 
components. Our results highlight the importance to consider the effect 
of ecological noise to predict the impact of seed banks on neutral and selective 
evolution.
\par\bigskip
\end{abstract}

\begin{keyword}
Diffusion \sep Moran model \sep seed bank \sep selection \sep site-frequency spectrum 
\MSC[2010] 92D10, 39A50,  60H10
\end{keyword}

\end{frontmatter}


\section{Introduction}

Genetic drift and natural selection are prominent forces shaping the amount of genetic 
diversity in populations. In diploid dioecious organisms, natural selection can be 
decomposed in different components: 1)~viability selection as the differential 
survival of the genotypes from zygotes to adults, 2)~fecundity (or fertility) 
selection as the differential zygote production, 3)~sexual selection as the differential 
success of the genotypes at mating, and 4)~gametic selection as the distorted segregation 
in heterozygotes~\citep{Bundgaard1972, Clegg1978}.
In effect, population genetic models with 
discrete generations or with Malthusian parameter ignoring age-structure lump these 
components into one unique selection parameter. Experimental or genomic population 
studies describing thus changes in allele frequencies often fail to describe 
and dissect the respective effects of these selective modes. Several theoretical 
studies on fertility~\citep{Bodmer1965} 
or on sexual selection~\citep{Karlin1969} 
as well as experimental work on animal~\citep{Prout1971a,Prout1971b,Christiansen1973}
and plant~\citep{Clegg1978}
populations have attempted 
to disentangle the respective influence of these selection components.
\\

In age-structured populations, however, genetic drift and selection 
can act differently than predicted by models without age structure 
(overview in the book by~\citealp{Charlesworth:book}).
The first type of age-structured models are simply obtained 
by individuals' life span and reproduction overlapping several 
generations. Here, genetic drift acts equally on all individuals 
of all age classes at any generation. The magnitude of genetic drift 
is defined by the population size which can be fixed, or fluctuating 
following a logistic dynamic size constrained by the population carrying 
capacity. As an extreme type of overlapping generation model, the Moran 
model exhibits a rate of random genetic drift all but the same as 
the Wright-Fisher (WF) model up to a rescaling of the population 
size~(see, e.g., the book by~\citealp[Chapter 7.2]{durrett}). Meanwhile, in age-structured 
populations, selection for fecundity and 
viability can show different outcomes, such as time to allele fixation and/or maintenance 
of alleles, compared to discrete models. This occurs if fecundity or longevity act 
at different ages of the structured population and under a logistic population size 
dynamic~\citep{Anderson1970,King1971,Charlesworth1972a,Charlesworth1972}. 
An interesting question 
arising from the current increasing availability of genomic data is whether 
in age-structured populations selection for fecundity can be disentangled from that 
of viability using population genomics statistics (such as the site-frequency 
spectrum, SFS).
\\

A second type of age-structured model is obtained when considering that individuals may 
remain as dormant/quiescent structures spanning several generations. Quiescence 
in reproductive structure is in fact wide-spread, such that seeds or eggs can be persistent 
states that allow to buffer a variable environment~\citep{evans2005}. 
The time at which offspring germinates or hedges can be variable, such that 
only some of the offspring live in detrimental periods, and most likely at least some in 
a beneficial environment. Bacterial spores or lysogenic states of temperate phages are also 
examples of such a bet-hedging strategy. Seedbanks represent thus a storage of genetic 
diversity decreasing the probability of population extinction~\citep{brown1977}, 
diminishing the effect of genetic drift~\citep{nunney2002}, 
slowing down the 
action of natural selection~\citep{Templeton1979,Koopmann2017} 
and favouring balancing selection~\citep{tellier2009}.
The strength of the seed bank effect clearly depends on the organism under consideration. 
Dormant seeds and diapausing eggs have a somehow short live span compared to 
the average coalescent time, 
and we call these seed banks ``weak''. Dormant states of bacteria, however, can last 
a longer time (many generations) even than the average coalescent time. These seed banks 
are called ``strong'', and modelled in  a similar way as weak selection: the time scale 
of the quiescent state is scaled by the inverse of the population size. In particular 
\citet{blath2015a,blath2015b} investigated in a series of papers strong 
seed banks, and find mathematically appealing results as deviations from the Kingman 
coalescent.
\\
In the present paper, we focus on weak seed banks aiming at applications to plant 
or invertebrate species. In a seminal paper \citet{kaj2001} 
investigated the effect of weak seed banks on the coalescent. 
The key parameter here is $G$ that denotes the average number of plant generations 
a seed rests in the soil. 
\citet{kaj2001}  
have obtained 
the Kingman $n$-coalescent rescaled by $(1+G)^{-2}$ (the nomenclature in the paper 
of \citet{kaj2001} is somewhat different, as they consider non-overlapping 
generations in a discrete time setting, while we formulate
the result already for a time continuous model with overlapping generations). 
It was subsequently shown 
that 
$1+G$ can be estimated using polymorphism data and information on the census size 
of populations~\citep{Tellier2011a}. 
Furthermore, neglecting seed banks may yield 
distorted results for the inference of past demography using for example the 
SFS~\citep{Zivkovic2012}. Interestingly, the effect 
of (weak) selection is enhanced by the slow-down of the time scale due to seed 
banks~\citep{blath2013,blath2015b,Koopmann2017}. The effect of genetic drift 
and 
weak natural selection on allele frequencies can be computed in 
a diffusion framework 
in a Moran model with deterministic seed bank ~\citep{Koopmann2017}. While 
the diffusion term, defining genetic drift is scaled by $(1+G)^{-2}$, matching 
the backward coalescent result of \citet{kaj2001}, the coefficient 
of natural selection in the drift term is multiplied only by $(1+G)^{-1}$. 
In biological 
terms, this means that the strength of selection, as defined by a unique 
selective 
coefficient, is enhanced by the seed bank compared to the effect 
of genetic drift, 
even though the time to reach fixation for an allele is 
increased~\citep{Koopmann2017}.
\par\medskip
We investigate two additions to this current body of theoretical literature on weak seed 
banks. 
First, we compute the effect of realistic models relaxing the hypotheses of a fixed size 
for the population above ground, and of deterministically large for the seed bank 
compartment. By doing so, we generate noise in the population size above or below ground. 
This extends the classic Moran or Wright-Fisher models, as the importance of the noise 
effect on population dynamics on the rate of genetic drift or selection is being 
recognized~\citep{Huang2015}. However, this approach is faced with an additional 
difficulty: if two alleles are present in a population of a constant size, 
it is sufficient to keep track of the number of individuals for one allele only. 
In the case of a fluctuating population size with logistic dynamics constrained 
by the carrying capacity, the state space becomes essentially two-dimensional. 
This more general case can still be examined via a time scale analysis and a dimension 
reduction by singular perturbation approaches. These methods are applicable 
in the case where the population size becomes 
large~\citep{Parsons2007,Parsons2008,kogan2014}. Though in those papers 
(as in the present one) the arguments are used in a formal way, the validity 
of this approach is proven~\citep[chapter 15.5 and quotations therein]{kuehn2015}. This latter analysis reveals 
that the dynamics are well described by appropriate scaled 
diffusion approximation of a Moran models.
Second, we dissect the fitness of plants into four components, which can be possibly 
affected by genetic drift occurring above ground and in the seed bank. We compute 
classic population genetics results for neutral and selected alleles and derive 
the expected SFS for the alleles under different fitness components. 
The analysis 
reveals the effect of noise in the above-ground population (plants) as well as 
in the below-ground population (seeds). In particular, we find that genetic drift 
and selection are differently influenced by above and below-ground noise. Additionally, 
the selection coefficient of alleles involved in seed death rate or germination rate 
are reduced by seed bank noise, but not by above-ground noise, while the two others 
are not affected (plant death rate and seed production). Below-ground noise also 
reduces the seed bank storage effect of neutral genetic diversity.

\section{Methods}
\begin{figure}
\begin{center}
\includegraphics[width=14cm]{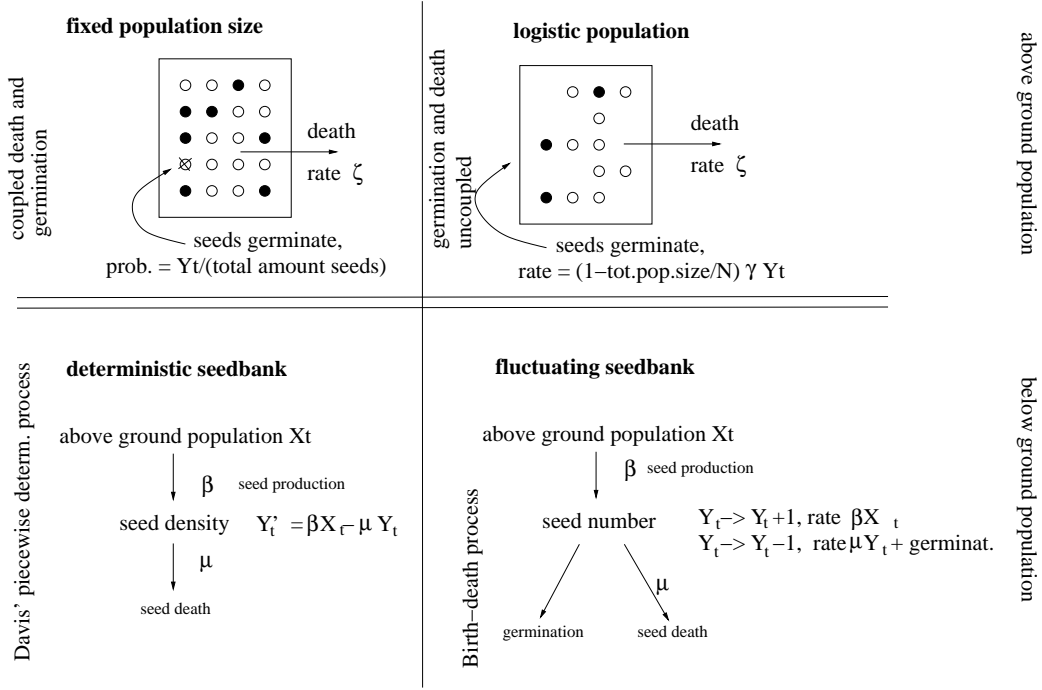}
\end{center}
\caption{Schematic sketch of the model ingredients: fixed population
size above ground (upper left), logistic population  model above ground (upper right), 
deterministic seedbank (lower left), fluctuating seedbank (lower right).}
\label{modelStruct}
\end{figure}

The aim of the present study is to investigate 
the effect of above-ground/below-ground population noise on evolution 
in presence of seed banks. We extend the results obtained 
in~\citep{Koopmann2017}, 
where a seedbank model with deterministic seedbank, fixed 
above ground population size and weak selection in seed survival 
and seed production 
has been considered. In the present paper, we consider more sources 
of variability and selection also in plant death rate and seed germination, 
and compare the results of four different models 
(Figure~\ref{modelStruct}):  fluctuations/no fluctuations in the total
above ground population, intrinsic fluctuations/no fluctuations in the seed bank.\\
Noise in population size can be located either 
above ground (plant population), or below ground (seed bank or seed population). 
Traditionally, the plant above-ground population is assumed
to consist of $N$ individuals, $N$ being fixed. A noisy substitute of this 
assumption is a logistic (fluctuating) population 
model~\citep{Anderson1970,King1971,Charlesworth1972a,Charlesworth1972}. 
For seeds, we find in 
the literature
either the assumption of a finite, fixed number~\citep{kaj2001} of seeds,
or, a deterministic infinite seed density~\citep{Koopmann2017}. 
In the latter, the assumption is that the number of seeds per plant 
is large enough, 
such that intrinsic noise is negligible. We suggest, here, an alternative 
model, 
in which each plant produces single seeds at time points that are 
distributed according 
to a Poisson process. The seeds also die (or loose their ability
to germinate) after a random time. 
Throughout the paper, we only consider Markovian processes, that is, 
all waiting times are exponentially distributed. 
The ``fluctuating  seedbank'' assumption assumes 
stochastically varying seed bank size. We therefore consider four models 
defined 
by all possible combinations of fixed or logistic above-ground population 
and deterministic or fluctuating seed bank. 
\par\medskip

{\bf Notation:} 
$X_{1,t}$ denotes the number of allele-$A$ plants. In a model with fixed above-ground 
population size $N$, the number of allele-$B$ plants reads $X_{2,t}=N-X_{1,t}$; 
in the logistic fluctuating version, we set up a stochastic process for $X_{1,t}$ 
and $X_{2,t}$. In that case, $N$ does denote the carrying capacity of the population 
(\textit{i.e.}, the maximal possible size). The average population size (conditioned on non-extinction) 
is $\kappa N$ for some $\kappa\in(0,1)$. 
$Y_t$ ($Z_t$) always refers to the amount of allele $A$ (allele $B$) seeds in the 
bank. For the deterministic seed bank, $Y_t$ ($Z_t$) are real 
numbers that follow (conditioned on $X_{1,t}$, $X_{2,t}$) an ordinary differential 
equation (ODE). In the case of stochastically fluctuating seed banks, 
$Y_t$ ($Z_t$) are non-negative integers, that follow a stochastic 
birth-death process.\par \medskip

We allow for weak natural selection. 
The rates for allele $B$ individuals slightly differ from those for allele 
$A$ individuals on a scale of $1/N$, as it is usual for weak 
effects. If $\sigma_i>0$, allele B has a disadvantage in comparison 
with allele A in the respective process (where $\sigma_1$
addresses the death of a plant, $\sigma_2$ the death of seeds, 
$\sigma_3$ the production of seeds, and $\sigma_4$ the germination of seeds). Of course, the signs 
of $\sigma_i$ can be chosen in an arbitrary way to consider 
a genotype $B$ that has an advantage above genotype A ($\sigma_i<0$), 
or for example a situation where B has a disadvantage 
above ground ($\sigma_1$, $\sigma_3>0$) and an 
advantage below ground ($\sigma_2$, $\sigma_4<0$). 
The parameters of our models are summarized in Table~\ref{table1}.\\

\begin{minipage}{\linewidth}
\begin{center}
\begin{tabular}{lll}
\hline\tabularnewline 
meaning & symbol (Allele $A$) & symbol (Allele $B$)\\
\tabularnewline\hline\tabularnewline 
death rate of  plants & $\zeta$ & $\zeta\,(1+\sigma_1/N)$\\
\tabularnewline death rate of seeds & $\mu$ & $\mu\,(1+\sigma_2/N)$\\
\tabularnewline production rate of seeds & $\beta$ & $\beta\,(1-\sigma_3/N)$\\
\tabularnewline germination rate of seeds (log.\ pop.\ only) & $\gamma$& $\gamma\,(1-\sigma_4/N)$\\
\tabularnewline\hline
\end{tabular}
\end{center}
\captionof{table}{Parameters of the models.}
\label{table1}
\end{minipage}\\[8pt]

\subsection{Fixed population size and deterministic seed bank}
\label{subsec:2.1}
This model has been developed and discussed before~\citep{Koopmann2017}. The total 
above-ground population has size $N$, and the transitions for the 
allele-$A$ plant population $X_{1,t}\in\{0,\ldots,N\}$ given the 
seed densities $Y_t$ and $Z_t\in\R_+$ are presented in Table~\ref{table2}. 
E.g., an A-plant dies at rate $\zeta X_{1,t}$. 
In the standard Moran model, it is instantaneously 
replaced by a B-individual with 
probability $1-X_{t,1}/N$. In our case, the seeds determine the probability 
for a B-individual, where this probability is given by $Z_t/(Y_t+Z_t)$. In the same way we obtain 
the rate at which a B-individual dies (the rate is $\zeta (1+\sigma_1/N)\,(N-X_{1,t})$) and is replaced by 
an A-individual (the probability is $Y_t/(Y_t+Z_t)$). \\

\begin{minipage}{\linewidth}
\begin{center}
\begin{tabular}{lll}
\hline\tabularnewline 
event & offset &rate\\
\tabularnewline\hline\tabularnewline 
death of $A$, birth of $B$&
$X_{1,t}\mapsto X_{1,t}-1$ &
$\zeta X_{1,t}\,\,Z_t/(Y_t+Z_t)$\\
\tabularnewline death of $B$, birth of $A$&
$X_{1,t}\mapsto X_{1,t}+1$ &
$\zeta (1+\sigma_1/N) (N-X_{1,t})\,\,Y_t/(Y_t+Z_t)$\\
\tabularnewline\hline
\end{tabular}
\end{center}
\captionof{table}{Possible transitions and their rates.}
\label{table2}
\end{minipage}\\[8pt]

In this model we assume that the number of seeds a plant produces 
is large (basically infinitely large), such that the seed density 
in the soil follows a deterministic process, given the history of 
the above ground population. The dynamics of seeds follows a Davis' piecewise 
deterministic process~\citep{Davis1984}, where plants produce seeds 
at rate $\beta$ (resp.\ $\beta(1-\sigma_3/N))$ and seeds die at rate 
$\mu$ (resp.\ $\mu(1+\sigma_4/N)$)
$$ 
\dot Y_t = \beta X_{1,t}-\mu Y_t,\quad
\dot Z_t = \beta(1-\sigma_3/N) (N-X_{1,t})-\mu(1+\sigma_2/N) Z_t.
$$

\subsection{Logistic population dynamics and deterministic seed bank}
\label{subsec:2.2}
For the logistic model, we do not couple death and birth of a plant as 
it is usually done in Moran-type models to keep the total population 
size constant. We generalize the logistic dynamics as investigated, e.g., 
by~\cite{Nasell2011}, or~\cite{Parsons2007,Parsons2008} 
for the situation at hand and separate birth and death events. If 
the seed densities $Y_t,Z_t\in\R_+$ are given, the transitions 
for $X_{1,t}$, $X_{2,t}\in\{0,\ldots,N\}$, $X_{1,t}+X_{2,t}\leq N$, 
read as summarized in Table~\ref{table3}.\\

\begin{minipage}{\linewidth}
\begin{center}
\begin{tabular}{lll}
\hline\tabularnewline 
 event & offset & rate \\
\tabularnewline\hline\tabularnewline  
death of $A$ & $X_{1,t}\rightarrow X_{1,t}-1$ & $\zeta X_{1,t}$ \\
\tabularnewline death of $B$ & $X_{2,t}\rightarrow X_{2,t}-1$ & $(1+\sigma_1/N)\zeta X_{2,t}$ \\
\tabularnewline birth of $A$ & $X_{1,t}\rightarrow X_{1,t}+1$ & $\gamma(1-(X_{1,t}+X_{2,t})/N)Y_t$ \\
\tabularnewline birth of $B$ & $X_{2,t}\rightarrow X_{2,t}+1$ & $(1-\sigma_4/N)\gamma(1-(X_{1,t}+X_{2,t})/N)Z_t$\\
\tabularnewline\hline
\end{tabular}
\end{center}
\captionof{table}{Possible transitions and their rates.}
\label{table3}
\end{minipage}\\[8pt]

Conditioned on $X_{1,t}$ and $X_{2,t}$, the dynamics of seeds are 
again deterministic, 
$$
\dot{Y} = \beta X_1 - \mu Y,\qquad
\dot{Z} = \left(1-\frac{\sigma_3}{N}\right)\beta X_2 - \left(1+\frac{\sigma_2}{N}\right)\mu Z.
$$

\subsection{Fixed population size and fluctuating seed bank}
\label{subsec:2.3}
For the above-ground population we return to a fixed population size, 
s.t.\ it is sufficient to follow $X_{1,t}$ as  $X_{2,t}=N-X_{1,t}$. 
In the present model we address the noise in the number of seeds, $Y_t,Z_t\in\N_0$. 
The seeds follow a stochastic birth-death process, where the death rate 
is kept constant, and the birth rate is proportional to the number of 
corresponding above-ground plants. We obtain the transitions summarized in Table~\ref{table4}.\\
\setlength{\tabcolsep}{6pt}

\begin{minipage}{\linewidth}
\begin{center}
\begin{tabular}{lll}
\hline\tabularnewline 
 event & offset & rate \\
\tabularnewline\hline\tabularnewline 
death of $A$, birth of $A$ & $(X_{1,t},Y_t)\rightarrow (X_{1,t},Y_t-1)$ & $\zeta X_{1,t} Y_t/(Y_t+Z_t)$ \\
\tabularnewline death of $A$, birth of $B$ & $(X_{1,t},Z_t)\rightarrow (X_{1,t}-1,Z_t-1)$ & $\zeta X_{1,t} Z_t/(Y_t+Z_t)$ \\
\tabularnewline death of $B$, birth of $A$ & $(X_{1,t},Y_t)\rightarrow (X_{1,t}+1,Y_t-1)$ & $(1+\sigma_1/N)\zeta (N-X_{1,t}) Y_t/(Y_t+Z_t)$ \\
\tabularnewline death of $B$, birth of $B$ & $(X_{1,t},Z_t)\rightarrow (X_{1,t},Z_t-1)$ &   $(1+\sigma_1/N)\zeta (N-X_{1,t}) Z_t/(Y_t+Z_t)$ \\
\tabularnewline birth of $A$-seed & $Y_t \rightarrow Y_t+1$ & $\beta X_{1,t}$ \\
\tabularnewline death of $A$-seed & $Y_t \rightarrow Y_t-1$ & $\mu Y_t$ \\
\tabularnewline birth of $B$-seed & $Z_t \rightarrow Z_t+1$ & $(1-\sigma_3/N)\beta (N-X_{1,t})$ \\
\tabularnewline death of $B$-seed & $Z_t \rightarrow Z_t-1$ & $(1+\sigma_2/N)\mu Z_t$ \\
\tabularnewline\hline
\end{tabular}
\end{center}
\captionof{table}{Possible transitions and their rates.}
\label{table4}
\end{minipage}\\[8pt]
\setlength{\tabcolsep}{10pt}

\subsection{Logistic population dynamics and fluctuating seed bank}
\label{subsec:2.4}

The last model incorporates logistic growth and a stochastically fluctuating 
seed bank. This model is an obvious combination of the last two models.\\

\begin{minipage}{\linewidth}
\begin{center}
\begin{tabular}{lll}
\hline\tabularnewline
 event & offset & rate\\
\tabularnewline\hline\tabularnewline
death of $A$ & $X_{1,t}\rightarrow X_{1,t}-1$ & $\zeta X_{1,t}$ \\
\tabularnewline death of $B$ & $X_{2,t}\rightarrow X_{2,t}-1$ & $(1+\sigma_1/N)\zeta X_{2,t}$ \\
\tabularnewline birth of $A$ & $(X_{1,t},Y_t)\rightarrow (X_{1,t}+1,Y_t-1)$ & $\gamma(1-(X_{1,t}+X_{2,t})/N)Y_t$\\
\tabularnewline birth of $B$ & $(X_{2,t},Z_t)\rightarrow (X_{2,t}+1,Z_t-1)$ & $(1-\sigma_4/N)\gamma(1-(X_{1,t}+X_{2,t})/N)Z_t$\\
\tabularnewline birth of $A$-seed & $Y_t \rightarrow Y_t+1$ & $\beta X_{1,t}$ \\
\tabularnewline death of $A$-seed & $Y_t \rightarrow Y_t-1$ & $\mu Y_t$ \\
\tabularnewline birth of $B$-seed & $Z_t \rightarrow Z_t+1$ & $(1-\sigma_3/N)\beta X_{2,t}$ \\
\tabularnewline death of $B$-seed & $Z_t \rightarrow Z_t-1$ & $(1+\sigma_2/N)\mu Z_t$\\
\tabularnewline\hline
\end{tabular}
\end{center}
\captionof{table}{Possible transitions and their rates.}
\label{table5}
\end{minipage}\\[8pt]

\subsection{Strategy for the analysis of the models}

The aim of the analysis is the reduction of the four models to 
a one-dimensional
diffusion approximation of a Moran model representing the fraction of allele-$A$ individuals within 
the population. As the details of the analysis are tedious, we present them in detail in the appendix and only outline the basic idea in the present section.\par\medskip

The strategy of the analysis differs for the first model 
(fixed population size, deterministic seed bank) and the other three models. 
The reason is that there is one single stochastic state 
variable $X_{1,t}$ in the first model, so that if we know the history of $X_{1,t}$, 
the state of the seed bank is known. No dimension 
reduction method is thus required. We basically can use the approach of~\cite{Koopmann2017} 
to derive the diffusion approximation of a Moran model. However, we outline in appendix~\ref{appendixModel1} 
a heuristic argument
based on a small-delay approximation, as this route seems to provide 
an appealing 
short-cut (though to our knowledge for this approximation, 
that is used in theoretical 
physics, no hard convergence theorem is available): if ecological 
time $t$ is not considered but rather the evolutionary 
time $\tau=t/N$ (population size $N$ large), the delay of 
a weak seed bank is small. In this case, 
the solution can be expanded w.r.t.\ the delay. 
As a result, the seed bank can be removed from 
the stochastic process and replaced by appropriately 
rescaled parameters. Since for the time beeing, the short delay approximation is only a heuristic approach, we formulate also a proof for the result based on time scale arguments (as explained next) in appendix~\ref{fixedPopoDetSeedII}.\par\medskip

\begin{figure}[h]
\centering
	\includegraphics[width=\textwidth]{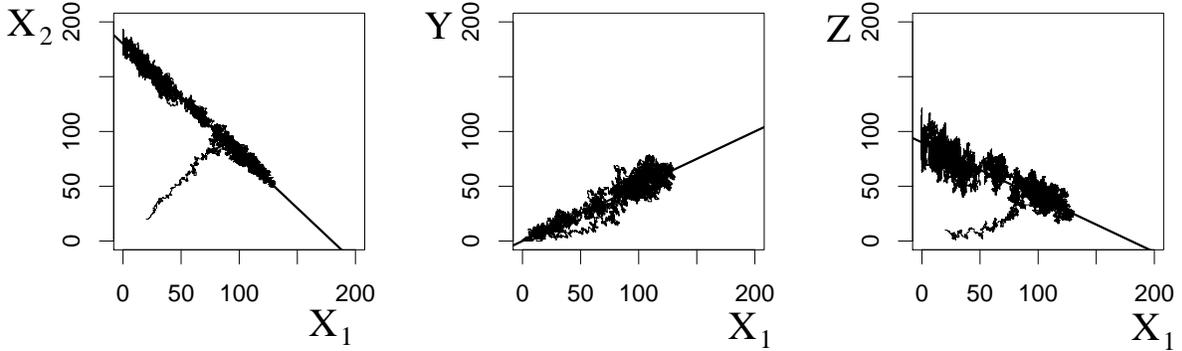}
\caption{Simulated trajectory for model~\ref{subsec:2.4}, 
(\textit{left}) $A$-plants $X_{1,t}$ vs $B$-plants $X_{2,t}$, 
(\textit{center}) $A$-plants $X_{1,t}$ vs $A$-seeds $Y_t$, 
(\textit{right}) $A$-plants $X_{1,t}$ vs $B$-seeds $Z_t$, $N=200$, 
$\beta  = 1$, 
$\mu = 1$, 
$\zeta = 0.5$, 
$\gamma  = 10$, $\sigma_i=0$, s.t.\ according to Proposition~\ref{propDetMod4}
$\theta=0.5$, and $\kappa=0.9$.
The solid line indicates the coexistence line in Proposition~\ref{propDetMod4}, 
scaled by the population size $N$.
}\label{SimLogNoise}
\end{figure}

The three other models have two, three or four stochastic state variables. 
Methods of dimension reduction are required to obtain a one-dimensional 
diffusion approximation of a 
Moran model. The key insight here is that any realization saddles fast 
on a one-dimensional manifold. If we consider, e.g., the deterministic version 
of the logistic population dynamics with stochastically fluctuating seed bank, 
we find, according to arguments by, e.g.,~\cite{Kurtz1980}, for the deterministic
limit as $N\rightarrow\infty$ 
(with $x_i(t)=X_{t,i}/N$, $y(t)=Y_t/N$, $z(t)=Z_t/N$) 
\begin{align*}
\dot{x_1} &= \gamma(1-x_1-x_2)y -\zeta x_1 \\
\dot{x_2} &= \gamma(1-x_1-x_2)z -\zeta x_2 \\
\dot{y} &= \beta x_1 - \mu y\\
\dot{z} &= \beta x_2 - \mu z 
\end{align*}
It is straightforward  to show that a line of stable equilibria, 
the so-called coexistence line, exists:
\begin{prop}\label{propDetMod4}
Let $\vartheta := (\beta-\zeta)/\mu>0$, $\kappa := (\gamma\vartheta-\zeta)/(\gamma\vartheta) \in [0,1]$. Then, there is a line of stationary points in $[0,\kappa]^2\times\mathbb{R}^2_+$ given by
\begin{align*}
(x_1,x_2,y,z)=(x,\kappa-x,\vartheta x,\vartheta(\kappa-x)), &&x\in[0,\kappa].
\end{align*}
The line of stationary points is transversally stable 
(locally and globally in the positive cone).
\end{prop}
It turns out that the stochastic process rapidly approaches  
this line of equilibria, and performs a random walk close to 
it (see Figure~\ref{SimLogNoise}). The analysis reveals that 
the distribution on a transversal cut is just a normal distribution with 
a variance of ${\cal O}(1/N)$. Along the line of stationary points, however, 
the realizations will move 
according to a one-dimensional diffusion approximation of a Moran process. 
This approximative process is a combination 
of one component of the full process parallel to this line, 
and a second component that results from an interaction between a component 
perpendicular to this line with the deterministic vector field directed 
towards this line. 
In order to reveal this structure, we first 
use a large population limit (Kramers-Moyal expansion)
to obtain a Fokker-Planck/Kolmogorov 
forward equation for the full process. In the second step we apply singular 
perturbation methods as described, e.g.,\ in~\cite{kogan2014} or~\cite[chapter 15.5]{kuehn2015} to perform the dimension 
reduction to the one-dimensional Moran model.

\section{Results}

\subsection{Timescales for different seed bank models}\label{ch:Timescales}

For all of our models, the resulting one-dimensional Fokker-Planck equation 
assumes the form of a diffusion approximation of a Moran-model with weak selection, 
\begin{equation}\label{eq:gen_Moran}
\partial_{\tau}u = -\sigma a\partial_{\x}\{\x(1-\x)u\} +\frac{1}{2}b\partial_{\x}^2\{\x(1-\x)u\}
\end{equation}
where $a$ and $b$ describe the speed of selection and genetic drift, respectively. The term 
$\sigma$ represents selective coefficients and is a generic parameter for notation including $\sigma_i$, $i=1,\ldots,4$. In order to formulate 
the results for $a$ and $b$, let us introduce three composite parameters: 
$G=\zeta/\mu$ is the number of plant generations a seed survives on average; 
$Y=\beta/\zeta$ is the average number of seeds produced by a plant; 
$\kappa$ already defined in Proposition~\ref{propDetMod4} is the average 
fraction of the 
above-ground population size in the logistic model, 
in comparison with the maximal possible 
population size $N$. Furthermore, for deterministic and fluctuating seed banks, we respectively denote $(1+G)^{-1}$ and $(1+(1-1/Y)G)^{-1}$ as $\mathcal{G}$, which can be seen as the number of plant generations that seeds survive on average corrected by the size of the seed bank. Using these abbreviations, the parameters $a$, $b$ 
and $\sigma$ for our four models are summarized in Table~\ref{table6}.\\

\begin{minipage}{\linewidth}
\begin{center}
\begin{tabular}[h]{llll}
\hline\tabularnewline
model/scale term & $a$(selection) & $b$(genetic drift) & $\sigma$ (selection coeff.)\\
\tabularnewline\hline\tabularnewline
fix. pop., det. s.b.  & $\zeta\;\mathcal{G}^{-1}$ & $2\zeta\;\mathcal{G}^{-2}$
& $\sigma_1+\sigma_2+\sigma_3$\\
\tabularnewline fix. pop., fluct.\ s.b. & $\zeta\;\mathcal{G}^{-1}$ & $2\zeta\;\mathcal{G}^{-2}$ 
&$\sigma_1+(1-1/Y)\sigma_2+\sigma_3$\\
\tabularnewline log.\ pop., det.\ s.b.  & $\zeta\;\mathcal{G}^{-1}$ & $2\zeta\;\mathcal{G}^{-2}\,\kappa^{-1}$ 
&$\sigma_1+\sigma_2+\sigma_3+\sigma_4$\\
\tabularnewline log.\ pop., fluct.\ s.b.  & $\zeta\;\mathcal{G}^{-1}$ & $2\zeta\;\mathcal{G}^{-2}\,\kappa^{-1}$ 
& $\sigma_1+\sigma_2+\sigma_3+(1-1/Y)\sigma_4$\\
\tabularnewline\hline
\end{tabular}
\end{center}
\captionof{table}{Drift, diffusion, and selection coefficients for the different population/seed bank models.}
\label{table6}
\end{minipage}\\[8pt]

The basic seed bank model (fixed population size, deterministic seed bank) 
and the standard Moran model without seed bank 
can be used as reference models. 
A seed bank slows down the time scale of selection as well as that of genetic drift \citep{Koopmann2017}, 
where selection is less affected (by a factor of $(1+G)^{-1}$) than genetic drift (by a factor of $(1+G)^{-2}$). 
We find that fluctuations in the above-ground population and in the seeds have different effects.\par\medskip
 Fluctuations in the seed number reduce the storage effect of seed banks. The additional noise yields a  
reduction of the effective time a seed spends within the seed bank, and thus increases the rate of genetic drift. For $Y\rightarrow \infty$ 
(noise in seed bank tends to zero), we obtain the result for the deterministic 
seed bank, for $Y\rightarrow 1$ (noise is maximized), the model converges towards 
the standard Moran model without seed bank. Note that $Y$ 
is not the average number of seeds per plant directly measured but the effective number of seeds per plants. 
For example, a certain fraction of seeds might be getting lost due to other environmental reasons (abiotic or biotic factors) than their intrinsic mortality. 
These seeds do not contribute to the bank. \par\medskip
The noise in the above-ground population only affects genetic drift and does not appear in the selection term. 
This result reflects that 
the actual competition between alleles $A$ and $B$ only happens above ground. 
Nonlinear terms in the transition rates only appear in the birth term of the plants. 
By increasing solely genetic drift, the above-ground noise can counteract the amplification of selection by seed banks.
\par\medskip
The scaling of selection by ${\mathcal G}^{-1}$ and that 
of  genetic drift by ${\mathcal G}^{-2}$ is somehow expected. 
All mutations are affected in the same 
way by the above-ground noise. Our result concerning the lumped 
selection coefficient $\sigma$, however, is unexpected: 
Mutations for some fitness components (mortality of seeds, $\sigma_2$, 
and germination ability, $\sigma_4$) show reduced selection while this is 
not the case for selective coefficients of other fitness components 
($\sigma_1$ and  $\sigma_3$). 
This means that if the number of seeds per plant is not too large, 
beneficial mutations in the mortality of seeds ($\sigma_2$) have 
a reduced chance to reach fixation compared with 
a beneficial mutation for the production of seeds ($\sigma_3$).

\subsection{Site-frequency spectrum (SFS)}

The SFS is a commonly used statistic for the analysis of genomewide distributed SNPs. It is defined as the distribution of the number of times a mutation is observed in a population or a sample of $n$ sequences conditional on segregation. 
Herein, this distribution is taken over numerous unlinked sites and mutations occur only on previously monomorphic ones \citep{Kimura1969} at rate $\theta/2$ per $N$ generations.  Each mutant allele that arises from the wildtype at such an independent site is assumed to marginally follow the diffusion model specified in (\ref{eq:gen_Moran}) so that in particular all mutants have equivalent selective effects among sites. Mutations are allowed to occur in plants and seeds, but the following results can be easily adapted to the scenario, where mutations may only arise in plants. The proportion of sites at equilibrium, where the mutant frequency is in $(y, y + dy)$, is routinely obtained as 
(e.g., \citealt{Griffiths2003,Koopmann2017})
\begin{equation}\label{eq:popSFS}
\hat{f}(y)=\frac{\theta}{b\; y (1-y)}\frac{1-\exp\{-2 a/b\;\sigma (1-y)\}}{1-\exp\{-2 a/b\;\sigma\}},
\end{equation}
where $a$, $b$ and $\sigma\neq{}0$ are given in Table~\ref{table6}. The sample SFS 
at equilibrium can be immediately obtained from (\ref{eq:popSFS}) via binomial 
sampling as 
\begin{equation}\label{eq:sampleSFS}
\hat{f}_{n,k}=\theta{}\frac{n}{b\; k(n-k)}\frac{1-{}_1F_1(k;n;2 a/b\;\sigma)e^{-2 a/b\;\sigma}}{1-e^{-2 a/b\;\sigma}},
\end{equation}
where $_1F_1$ denotes the confluent hypergeometric function of the first 
kind \citep{abramowitzandstegun}. The neutral versions 
of (\ref{eq:popSFS})~and~(\ref{eq:sampleSFS}) are respectively given 
by $\hat{f}(y)=\theta/(b y)$ and $\hat{f}_{n,k}=\theta/(b k)$. \par\medskip

\begin{figure}[h!]
	\centering
		\includegraphics[width=0.45\textwidth]{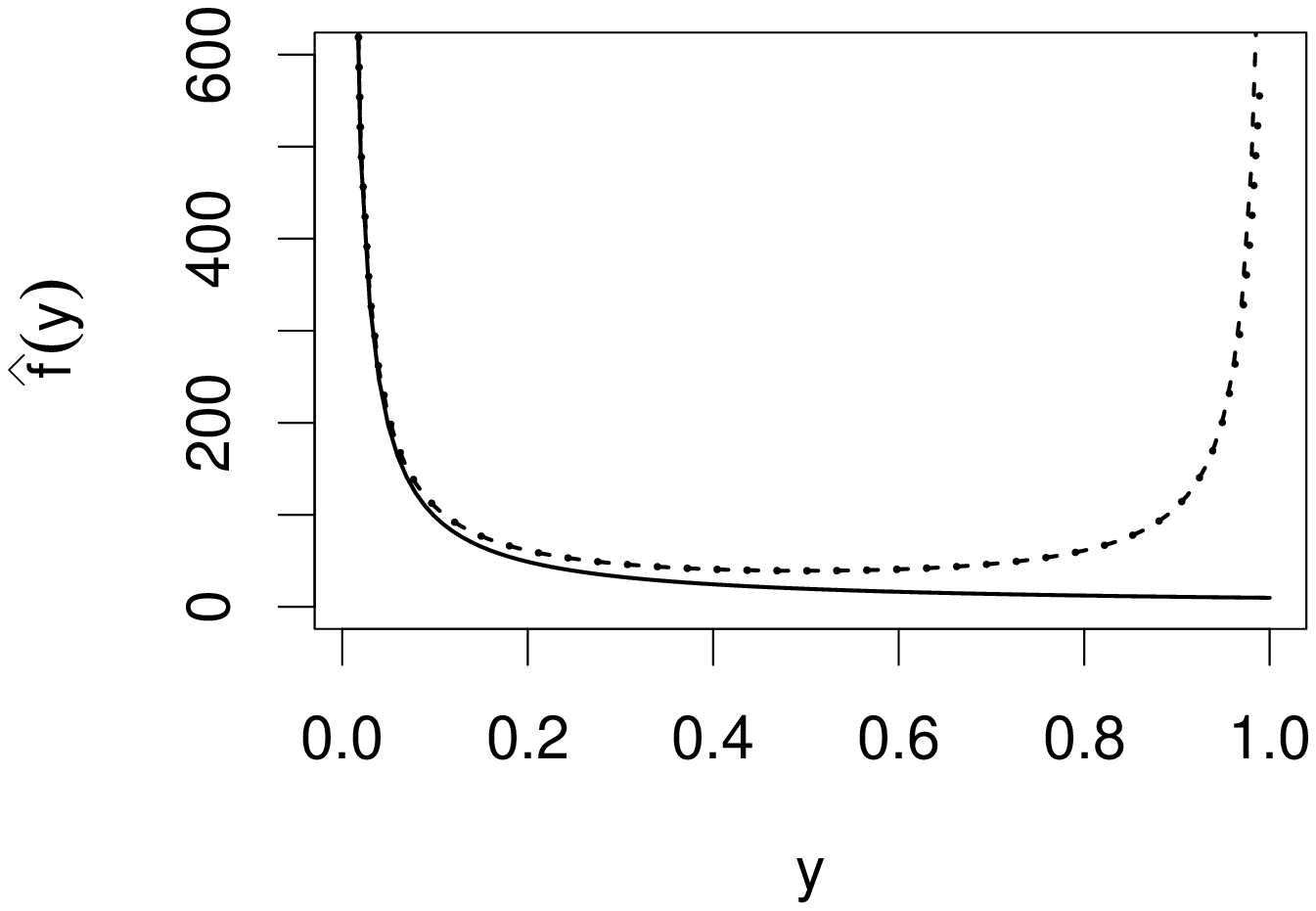}
		\includegraphics[width=0.45\textwidth]{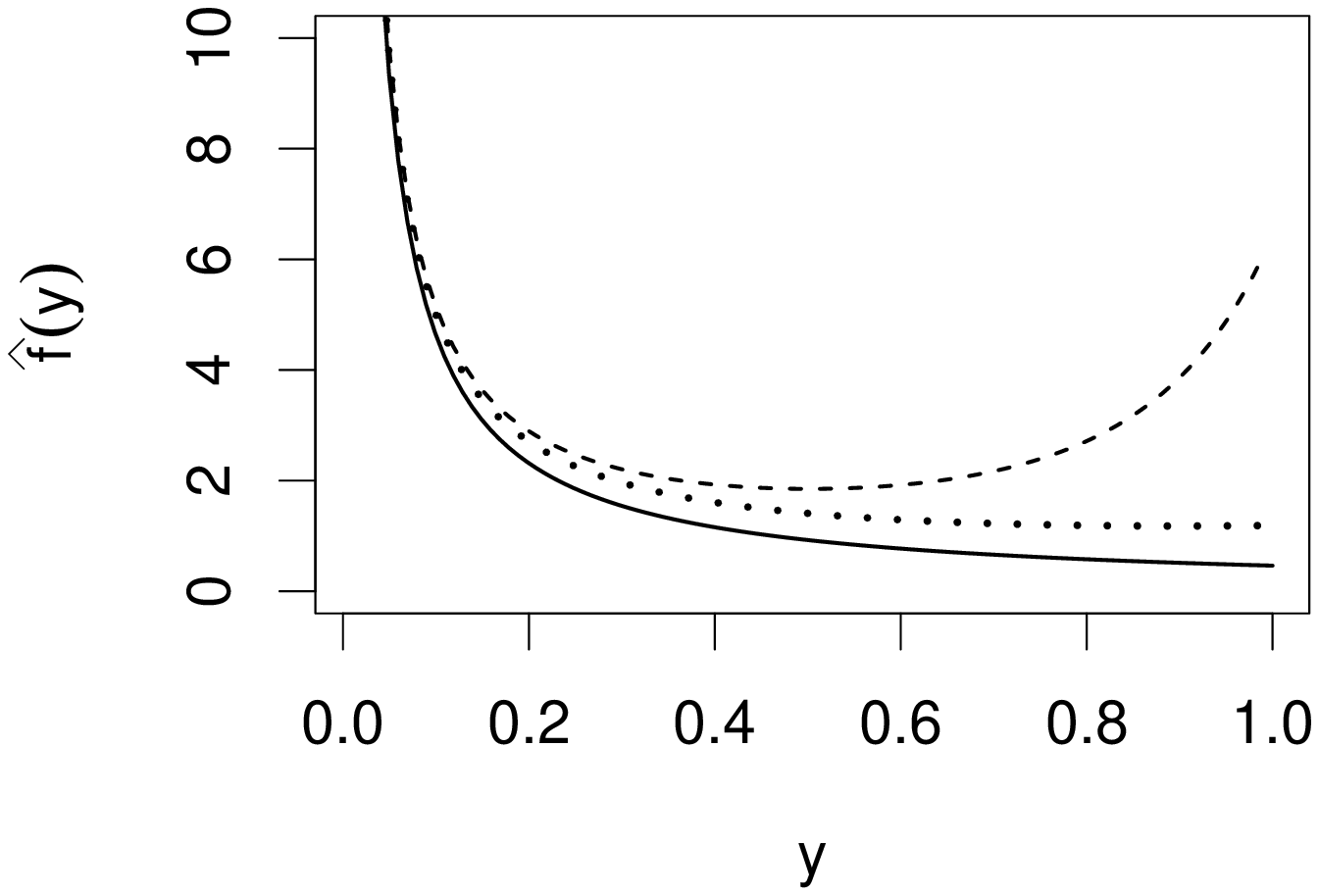}
	\caption{Continuous SFS $\hat f(y)$ over mutant frequency $y$ 
	according to (\ref{eq:popSFS}); neutral (solid) vs. selection 
	  (dashed, $\sigma_1=50$, $\sigma_2=\sigma_3=\sigma_4=0$, selection in the death rate of plants) 
resp.\ (dotted, $\sigma_1=\sigma_2=\sigma_3=0$, $\sigma_4=50$, selection in the germination rate of seeds).
	Further parameters: $\mu = 0.25$, $\zeta = 1$, $\gamma = 1.5$, $\theta=2$; 
	(left)  $\beta  = 2.4$, s.t.\ $G=4$ and $Y=2.4$, 
	(right)  $\beta  = 1.2$, s.t.\ $G=4$ and $Y=1.2$.	Note that in the 
	left panel the dashed and the dotted curve are almost on top 
	of each other.
	}
	\label{fig:SFS_1}
\end{figure}

Fig.~\ref{fig:SFS_1} indicates the striking effects of the seed bank noise 
for the model with logistic population dynamics and fluctuating seed banks. 
When the noise is low ($Y$ high) the number of segregating sites is expected 
to be high, and mutations involved in the selection coefficients $\sigma_1$ 
and $\sigma_4$ lead to similar SFS. The SFS show the typical U-shape expected 
under positive pervasive selection (left panel). If, however, $Y$ becomes small, the number 
of segregating sites decreases, and the selection coefficient  $\sigma_1$ shows 
a U-shaped SFS while mutations under $\sigma_4$ do not (right panel).

\section{Discussion}

We considered four models to investigate the effect of combined noise 
in the above-ground population and below-ground seed bank. In all
four cases time scale arguments allow us a reduction to 
a diffusion approximation of a Moran model. Our results extend 
the findings that a seed bank without noise yields a change of 
time scale in selection and genetic
drift that amplifies the effect of weak selection~\citep{Koopmann2017}. 
The first main result of this present work is that there is no direct 
interaction between the noise above ground and below ground. 
The above-ground noise increases the effect of genetic drift 
compared to a fixed above-ground population size, but does 
not affect selection. One can propose a new definition of 
the effective population size $N_e =  \kappa\,{\mathcal G}^2\,N/2$ 
describing the change in genetic drift due to this noise 
(according to~\citet[Definition 2.9]{etheridge}. 
This allows us to redefine the evolutionary time scale 
$T = \zeta\,t/N_e$. 
As a result of this procedure the parameter $\kappa$, 
which represents the reduction of the average 
total population size and indicates the increase of the above-ground 
noise by the logistic population dynamics, appears as a factor 
in front of the selection 
term in the Moran model. Note that in this notation
we find that the terms for below-ground 
(${\mathcal G}^2$) and above-ground ($\kappa$) noise 
are multiplied indicating their independence. \par\medskip

The second result is that the below-ground noise affects the scaling of time 
for both the selection and the 
genetic drift term. We introduce the concept of the mean effective number of 
seeds per plant in a similar definition to the effective population 
size~\citep{Wright1931}. 
If this average seed number tends to infinity, we recover the effect of 
a deterministic seed bank (no noise), and if this number tends to one, 
the noise of the seed bank accelerates the time scale such that 
the seed bank has no effect at all. The magnitude of noise in the seed bank is 
thus tuned between the two extreme cases of ``no seed bank'' (minimal effect) 
and ``deterministic seed bank'' (maximal effect). 
\par\medskip

The third result of the present study is 
the insight that below-ground noise may affect the four plants' fitness components 
of viability and fecundity differentially. If the 
below-ground noise is large, the selective effect of 
mutations involved in seed death and germination may even be cancelled out. 
Other mutations involved in fitness traits of the above-ground population, such 
as plant death or seed production are not affected. In other words, while finite 
size and noise above ground do not affect selection, noise and finite size of the 
seed bank do change the selection coefficients.
In biological terms we interpret this result as follows. 
Above ground, the fate of an allele under positive selection 
is classically determined by the strength of genetic drift, which 
depends on the population size (the diffusion term in the Fokker-Planck equation) 
compared to the strength of selection which depends on the selection 
coefficient (the drift term in the forward model). Under a deterministic 
seed bank (as in the present models and in~\citet{Koopmann2017}), selection 
is efficient because it occurs on plants, when they are above ground, with a probability $(1+G)^{-1}$ and genetic drift occurs on a coalescent scale of $(1+G)^{-2}$. 
Any change in the allele frequencies above ground translates directly into the deterministic 
seed bank, just with a small time delay. However, when the seed bank has 
a fluctuating finite size, the strength of selection on seed fitness (the 
coefficients $\sigma_2$ and $\sigma_4$) is decreased by the noise in the 
seed compartment. This occurs because selection and genetic drift in the bank 
occur at every generation and not only when seeds germinate.\par\medskip

This observation implies that competition experiments, which measure and 
compare the effect of mutations and/or allow to determine selection coefficients 
on the ecological time scale can hardly be used to extrapolate to the evolutionary 
time scale. Indeed, even if two mutations seem to have an equivalent value for 
the plant fitness in a competition experiment, the presence of seed bank noise may 
lead to different evolutionary outcomes.\par\medskip

We finally discuss two ways of using genome polymorphism data to estimate seed bank 
parameters as well as the selection coefficients. 
We can attempt to infer the seed bank parameters based on neutral genetic diversity under the idealized conditions that the sample SFS reached an equilibrium as $\hat{f}_{n,k}$ (\textit{i.e.}, there is no recent demographic impact) and that we can measure or estimate the population mutation rate $\theta$ and the death rate of plants $\zeta$. It turns out that $\mathcal{G}$ (the number of plant generations a seed survives on average) and $\kappa$ (the average above-ground 
population size) can then be identified, so that it is possible to disentangle the effects 
of the seed bank from those of the above-ground noise on neutral evolution by utilizing $\hat{f}_{n,k}$. However, 
we cannot identify the below-ground noise, as $G$ 
and $Y$ only appear in the composite parameter $\mathcal{G}$. Note that if we only have information about the relative 
sample SFS $\hat{f}_{n,k}/\sum_i\hat{f}_{n,i}$, the multiplicative 
constant in $f_{n,\cdot}$ cancels out, and therefore only $\kappa\,\mathcal{G}$ and the combined effect of above- and below-ground noise can be estimated. 
Extending the work by~\cite{Tellier2011a}, 
we suggest that the number of plant 
generations a seed survives on average can be estimated from the absolute SFS using for example a Bayesian inference method with 
priors on the census size of the above-ground population and the death rate of plants.\par\medskip

We may also aim to infer the selection coefficients underpinning the various 
plants' fitness components, namely the fecundity and viability of plants or seeds. 
With the abundance of gene expression, molecular and Gene Ontology data, it becomes feasible 
to group genes by categories of function or pathways, for example to know the genes involved in seed 
germination or seed integrity, viability and seed dormancy (e.g.,~\citealp{Righetti2015}). These functional groups of genes actually underlie the different plants' fitness components investigated in this study. 
Using genome-wide polymorphism data of several individuals, the SFS for each functional groups of genes can be computed and used to draw inference of selection, for example as the distribution of selective effects (e.g., the method by~\citealp{Eyre-Walker2009}). Our prediction is thus that the SFS would reflect the differential selection on these 
fitness components and can be observed over the genes involved in the different functional groups. The limitation in current data lies so far on the functional side, as more gene expression study are needed to assigne genes to functional network and to different plant fitness components.
As an extension, our expected SFS shows that the behavior of the seed fitness coefficients can be affected 
by the below-ground noise. So, we predict that populations with a small sized seed bank 
should exhibit less selection signatures on genes related 
to seed fitness compared to populations with a larger seed bank compartment. The same analysis as above can be conducted, but now comparing the SFS and inferred selection for the different functional groups of genes across several populations with know ecological set-ups which define the seed bank size. We further suggest that such procedure can be applied to disentangle selection for fecundity from that for viability in age-structured populations, if functional groups of genes can be assigned to these traits (for example using gene expression at different life stages and ages).

Our results differ thus from those of classic age-structured populations by the overlap 
of generations, as the seed bank can present its own rate of genetic drift. Moreover, 
selection acts differently above ground and below ground on the different plants' fitness 
components, which may allow us to disentangle their effect on the overall selection coefficient.

\section*{Acknowledgements}
{\it This research is supported in part by Deutsche Forschungsgemeinschaft grants TE 809/1 (AT), STE 325/14 from the Priority Program 1590 (DZ), 
and MU2339/2-2 from the Priority Program 1617 (JM).}

\section*{References}
\bibliographystyle{elsarticle-harv}
\bibliography{popGen}

\setlength{\tabcolsep}{12pt}

\newpage
\begin{appendix}
\renewcommand{\thesection}{Appendix}
\section{Analysis of the Models}
\renewcommand\thesection{\Alph{section}}

We present the conceptions for the analysis of the model of a
fixed population size and a deterministic seed bank as well as for 
logistic population dynamics and a deterministic seed bank in detail. Note that the computations for the remaining two models  
are similar but even more extensive, so that we do not present them in full length but only mention the results of the 
main steps. The computations for the dimension reduction have been checked using 
the computer algebra package MAXIMA~\citep{maxima} (see the available supplementary files).

\subsection{Fixed population size and deterministic seed bank}
\label{appendixModel1}
To keep the demonstration short, we present a nice  
but heuristic argument using the 
idea of a short delay approximation~\citep{Guillouzic1999}.
To our knowledge no rigorous 
approximation theorem is available, therefore we 
also give an alternative approach, based on time scale
analysis, at the end of the appendix (appendix~\ref{fixedPopoDetSeedII}). Both approaches yield
identical results. \par\medskip

Using the variation-of constant formula, we find 
$$
Y_t = \beta\int_0^\infty e^{-\mu s}X_{t-s}\, ds,
\qquad
Z_t = \beta(1-\sigma_3/N)\int_0^\infty e^{-\mu(1+\sigma_2/N)s}(N-X_{t-s})\, ds.
$$
Let $x_t=X_{1,t}/N$ and $\varepsilon^2=1/N$. Then, 
$$X_{1,t}\rightarrow X_{1,t}+1\mbox{ at rate }
\,\,\frac{\varepsilon^{-2}(1+\sigma_1\varepsilon^2)\,\zeta (1-x_t)\,\int_0^\infty e^{-\mu s}x_{t-s}\, ds}{\int_0^\infty e^{-\mu s}x_{t-s}\, ds+(1-\sigma_3\,\varepsilon^2)\int_0^\infty e^{-\mu(1+\sigma_2\,\varepsilon^2)s}(1-x_{t-s})\, ds},
$$
$$X_{1,t}\rightarrow X_{1,t}-1\mbox{ at rate }
\,\,\frac{\varepsilon^{-2}\,\zeta x_t\,(1-\sigma_3\eps^2) \,\int_0^\infty e^{-\mu(1+\sigma_2\,\varepsilon^2)(t-s)}(1-x_{t-s})\, ds}{\int_0^\infty e^{-\mu s}x_{t-s}\, ds+(1-\sigma_3\,\varepsilon^2)\int_0^\infty e^{-\mu(1+\sigma_2\,\varepsilon^2)s}(1-x_{t-s})\, ds}.
$$
With standard arguments, we obtain a stochastic 
delay differential equation (SDDE) at the evolutionary 
time scale $\tau = t\,\varepsilon^2$ as
\begin{eqnarray*}
dx_\tau &=& 
\bigg(\frac{\varepsilon^{-4}(1+\sigma_1\varepsilon^2)\,\zeta (1-x_\tau)\,\int_0^\infty e^{-\mu s/\eps^2}x_{\tau-s}\, ds}{\eps^{-2}\int_0^\infty e^{-\mu s/\eps^2}x_{\tau-s}\, ds+\eps^{-2}(1-\sigma_3\,\varepsilon^2)\int_0^\infty e^{-\mu(1+\sigma_2\,\varepsilon^2) s/\eps^2}(1-x_{\tau-s})\, ds}\\
&&-
\frac{\varepsilon^{-4}\,\zeta x_\tau\,\int_0^\infty e^{-\mu(1+\sigma_2\,\varepsilon^2)s/\eps^2}(1-x_{\tau-s})\, ds}{\eps^{-2}\int_0^\infty e^{-\mu s/\eps^2}x_{\tau-s}\, ds+\eps^{-2}(1-\sigma_3\,\varepsilon^2)\int_0^\infty e^{-\mu(1+\sigma_2\,\varepsilon^2)s/\eps^2}(1-x_{\tau-s})\, ds}
\bigg)\, d\tau\\
&+& \bigg(\frac{\eps^{-2}\,(1+\sigma_1\varepsilon^2)
\,\zeta (1-x_\tau)\,\int_0^\infty e^{-\mu s
/\eps^2}x_{\tau-s}\, ds}{\eps^{-2}\int_0^\infty 
e^{-\mu s/\eps^2}x_{\tau-s}\, ds+\eps^{-2}
(1-\sigma_3\,\varepsilon^2)\int_0^\infty 
e^{-\mu(1+\sigma_2\,\varepsilon^2)s/\eps^2}
(1-x_{\tau-s})\, ds}\\
&&+
\frac{\eps^{-2}\,\zeta x_\tau\,\int_0^\infty e^{-\mu(1+\sigma_2\,\varepsilon^2)s/\eps^2}(1-x_{\tau-s})\, ds}{\eps^{-2}\int_0^\infty e^{-\mu s/\eps^2}x_{\tau-s}\, ds+\eps^{-2}(1-\sigma_3\,\varepsilon^2)\int_0^\infty e^{-\mu(1+\sigma_2\,\varepsilon^2)s/\eps^2}(1-x_{\tau-s})\, ds}\bigg)^{1/2}\,dW_\tau.
\end{eqnarray*}
We aim a small delay approximation. Therefore, we note that for a function $\Phi(t)$, which is sufficiently smooth and bounded, we have (for $\mu>0$)
\begin{eqnarray*} 
\eps^{-2}\,\mu\,\int_0^\infty e^{-\mu(\tau-s)/\eps^2}\Phi(\tau-s)\, ds
&=& \eps^{-2}\int_0^\infty \mu\,e^{-\mu s/\eps^2}(\Phi(\tau)-s\Phi'(\tau)+{\cal O}(s^2))\, ds\\
&=& \Phi(\tau)-\eps^2\mu^{-1}\Phi'(\tau)+{\cal O}(\eps^4).
\end{eqnarray*}
Thus, at a formal level, 
$$\lim_{\eps\rightarrow 0}  \zeta\left(\eps^{-4}\,\mu\,\int_0^\infty e^{-\mu s/\eps^2}x_{\tau-s}\, ds-\eps^{-2}\,x_{\tau}\right)d\tau
=
-\frac \zeta \mu\, dx_\tau
=
-G\, dx_\tau,
$$
where $G=\zeta/\mu$. 
Note that this equation has to be interpreted in 
terms of the Euler-Maruyama approximation of an SDDE, 
where differential quotients are replaced by difference quotients. 
We again emphasize that this approach is only 
mend to be formal, as to our knowledge no rigorous approximation theorems 
for small delay approximations in the context of SDDE are available. The result is 
consistent with the time scale analysis of the 
present model (appendix~\ref{fixedPopoDetSeedII}); we conjecture that the short delay approach yields a valid approximation under suited (rather general) conditions. 
If we add 
$$\frac{-\,\zeta\left(\eps^{-4}\,\mu\,\int_0^\infty e^{-\mu s/\eps^2}x_{\tau-s}\, ds-\,\eps^{-2}\,x_{\tau}\right)d\tau}
{\eps^{-2}\mu\,\int_0^\infty e^{-\mu s/\eps^2}x_{\tau-s}\, ds+(1-\sigma_3\,\varepsilon^2)\,\eps^{-2}\,\mu\,\int_0^\infty e^{-\mu(1+\sigma_2\,\varepsilon^2)s/\eps^2}(1-x_{\tau-s})\, ds}
$$ 
to both sides of the SDDE and let $\eps\rightarrow 0$, we obtain
\begin{eqnarray*}
(1+G)dx_\tau &=& 
\zeta(\sigma_1+\sigma_2+\sigma_3)\, x_\tau(1-x_\tau)\, d\tau
+
(2\zeta x_\tau(1-x_\tau))^{1/2}\, dW_\tau.
\end{eqnarray*}
This equation yields the desired result with 
$\sigma=\sigma_1+\sigma_2+\sigma_3$.

\subsection{Logistic population dynamics and deterministic seed bank}
\label{appendixModel2}
Let $p_{i,j}(k,l,t)=\mathbb{P}(X_{1,t}=i,X_{2,t}=j,Y_t\in(k,k+dk),Z_t\in(l,l+dl))$ be the 
joint probability density of the resulting stochastic process (with discrete $i,j$ and continuous $k,l$). The corresponding master equation reads:
\begin{equation}
\begin{split}
&\dot{p}_{i,j}(k,l,t) + \nabla \left[\begin{pmatrix}\beta i-\mu k\\ 
\beta\,(1-\sigma_3/N) j-\mu\,(1+\sigma_2/N) l\end{pmatrix}p_{i,j}(k,l,t)\right] \\
= &-\left[\zeta (i + (1+\frac{\sigma_1}{N})j) + \gamma\frac{N-i-j}{N}(k+(1-\frac{\sigma_4}{N})l\right]p_{i,j}(k,l,t) \\ 
&+ \zeta(i+1)p_{i+1,j}(k,l,t) + (1+\frac{\sigma_1}{N})\zeta(j+1)p_{i,j+1}(k,l,t) \\ 
&+ \gamma\frac{(N-i-j+1)k}{N} p_{i-1,j}(k,l,t) + (1-\frac{\sigma_4}{N})\gamma\frac{(N-i-j+1)l}{N} p_{i,j-1}(k,l,t),\\
\end{split}
\end{equation}
where the operator $\nabla$ acts with respect to continuous state space variables $k,l$. 
Standard arguments yield the Fokker-Plank-approximation 
for large populations, where \mbox{$x_1=i/N$, $x_2=j/N$, $y=k/N$ and $z=l/N$, as} 
\begin{equation}\label{eq:FP_model2}
\begin{aligned} 
&\quad\d_tu(x_1,x_2,y,z,t) \\
&=\partial_{x_1}\left\{\bigg[\zeta x_1-\gamma(1-x_1-x_2)y\bigg]u(x_1,x_2,y,z,t)\right\} \\
&\quad+\partial_{x_2}\left\{\bigg[(1+\frac{\sigma_1}{N})\zeta x_2-(1-\frac{\sigma_4}{N})\gamma(1-x_1-x_2)z\bigg]u(x_1,x_2,y,z,t)\right\} \\
&\quad+\d_y\left\{\bigg[(\mu y -\beta x_1)\bigg]u(x_1,x_2,y,z,t)\right\} \\
&\quad+\d_z\left\{\bigg[(1+\frac{\sigma_2}{N})\mu z -(1-\frac{\sigma_3}{N})\beta x_2\bigg]u(x_1,x_2,y,z,t)\right\} \\
&\quad+\frac{1}{2N}\d_{x_1}^2\left\{\bigg[\zeta x_1+\gamma(1-x_1-x_2)y\bigg]u(x_1,x_2,y,z,t)\right\} \\
&\quad+\frac{1}{2N}\d_{x_2}^2\left\{\bigg[(1+\frac{\sigma_1}{N})\zeta x_2+(1-\frac{\sigma_4}{N})\gamma(1-x_1-x_2)z\bigg]u(x_1,x_2,y,z,t)\right\}.\\
\end{aligned}
\end{equation}
Note that the second order noise terms are solely 
due to $x_1$ and $x_2$; no noise is added by the 
seed bank variables $y$ and $z$. 

\subsubsection{Deterministic model}
The corresponding deterministic model 
(drift terms only)  yields the ODEs
\begin{align*}
\dot{x_1} &= \gamma(1-x_1-x_2)y -\zeta x_1, \\
\dot{x_2} &= \gamma(1-x_1-x_2)z -\zeta x_2, \\
\dot{y} &= \beta x_1 - \mu y,\\
\dot{z} &= \beta x_2 - \mu z. 
\end{align*}
The lifetime reproductive success (or basic reproduction number) of a plant reads 
$R_0=\beta\,\gamma/(\mu\zeta)$.
If $R_0>1$, the plant population can persist.
Under this condition, there is a line of stationary solutions:
\begin{prop}\label{DetLogPopDetSeed}
Assume $R_0>0$. 
Let $\vartheta := \beta/\mu$, $\kappa := (\gamma\vartheta-\zeta)/\gamma\vartheta$. 
 Then, $\kappa\in[0,1]$, and there is a line of stationary points in $[0,\kappa]^2\times\mathbb{R}^2_+$ given by
\begin{align*}
(x_1,x_2,y,z)=(\kappa\,x,\kappa\,(1-x),\vartheta \,\kappa\,x,\vartheta\,\kappa\,(1-x)), &&x\in[0,1].
\end{align*}
The line of stationary points is transversally stable (locally and globally).
\end{prop}
\textbf{Proof:} 
It is straightforward to check that the line indicated above 
consists indeed of stationary points with non-negative values. The local stability is a consequence of Hartman-Grobman 
and the analysis of the Jacobian (due to the block structure 
of this matrix the eigenvalues can be stated explicitly). 
Note that one eigenvalue necessarily is zero with an eigenvector pointing in the direction of the line of stationary points. \par
Now, to show global stability, we prove that the system will approach the equilibrium line from any starting point. We first respectively denote $P=x_1+x_2$ and $S=y+z$ as the total plant and seed populations and consider the resulting reduced system
$$\dot{P} = \gamma(1-P)S -\zeta P, \quad 
\dot{S} = \beta P - \mu S.
$$
Here we use that we only consider weak selection: all 
selection effects tend to zero for $N\rightarrow\infty$, and hence the ODE describes the neutral case. The divergence 
of this system is negative, and thus the combination of the 
theorems of Bendixon-Dulac and Pointcar\'e-Bendixon imply 
that trajectories $(P,S)$ tend to stationary points. 
This observation yields the desired global stability.
\qed\par\medskip

Note that in equilibrium, $x_1+x_2=\kappa$. That is, $\kappa\, N$ 
represents the average above-ground population size of the model 
conditioned on non-extinction.\par\medskip

\subsubsection{Dimension reduction by time scale analysis}
The computations in this section follow 
closely the calculations in the paper of~\cite{kogan2014} to perform a dimension reduction
by a time scale analysis. First, new local variables for the boundary layer around the equilibrium line are defined as
\begin{eqnarray}
x_1&=&\kappa\,\x +\frac  \varepsilon 2 \,\vt, 
\,\,\, x_2 = \kappa\,(1-\x) +\frac  \varepsilon 2 \,\vt, \\
\,\,\, y &=& \kappa\,\vartheta\,\x+\frac  \varepsilon 2\,(\y+\z+\vartheta\,\vt), 
\,\,\, z =  \kappa\,\vartheta\,(1-\x) +\frac \varepsilon 2 \,(\y-\z+\vartheta\,\vt),\nonumber
\end{eqnarray}
where
\begin{equation*}
\x=\frac{x_1-x_2+\kappa}{2\kappa}, \quad \vt = \varepsilon^{-1}(x_1+x_2-\kappa), \quad \y = \varepsilon^{-1}(y+z-\vartheta(x_1+x_2)), \quad \z = \varepsilon^{-1}(y-z-\vartheta(x_1-x_2),
\end{equation*}
and $\varepsilon^2=1/N$. 
For the transformed density $\rho(\x,\vt,\y,\z,t;\varepsilon)=u(x_1,x_2,y,z,t)$, we find 
\begin{equation}\label{eq:FP_re_model2}
\d_t\rho(\x,\vt,\y,\z,t;\varepsilon) = \left(\L+\varepsilon\LL + \varepsilon^2L^{(2)}\right)\rho(\x,\vt,\y,\z,t) + \mathcal{O}(\varepsilon^3),
\end{equation}
with linear differential operators
%
\begin{align}
L^{(0)}\rho =\:&\d_{\vt}\bigg\{\bigg[\frac{-\zeta}{\vartheta}\y+\gamma\vartheta\kappa\vt\bigg]\rho\bigg\}+\d_{\y}\bigg\{\bigg[(\mu+\zeta)\y-\gamma\vartheta^2\kappa\vt\bigg]\rho\bigg\} \nonumber\\
&+\d_{\z}\left\{\bigg[(\mu+\zeta)\z+\gamma\vartheta^2\kappa(1-2\x)\vt\bigg]\rho\right\} \nonumber\\
&+\d_{\vt}^2\bigg\{\zeta\kappa\rho\bigg\} +\d_{\y}^2\bigg\{\zeta\vartheta^2\kappa\rho\bigg\} + \d_{\z}^2\bigg\{\zeta\vartheta^2\kappa\rho\bigg\} \nonumber\\
&+ \d_{\vt}\d_{\y}\bigg\{-2\zeta\vartheta\kappa\rho\bigg\} + \d_{\vt}\d_{\z}\bigg\{2\zeta\vartheta\kappa(1-2\x)\rho\bigg\} + \d_{\y}\d_{\z}\bigg\{-2\zeta\vartheta^2\kappa(1-2\x)\rho\bigg\},\\
L^{(1)}\rho=\:& \d_{\x}\left\{\left[\frac{-\zeta}{2\vartheta\kappa}\z-\frac{\gamma\vartheta}{2}(1-2\x)\vt\right]\rho\right\}+\d_{\vt}\left\{\bigg[\gamma\vt\y+\gamma\vartheta\vt^2+(\sigma_4+\sigma_1)\zeta\kappa(1-\x)\bigg]\right\}\nonumber\\
&+\d_{\y}\left\{\bigg[-\gamma\vartheta\vt\y-\gamma\vartheta^2\vt^2-(\sigma_1+\sigma_4)\zeta\vartheta\kappa(1-\x)+(\sigma_3+\sigma_2)\beta\kappa(1-\x)\bigg]\rho\right\}\nonumber\\
&+\d_{\z}\left\{\bigg[-\gamma\vartheta\vt\z+(\sigma_4+\sigma_1)\zeta\vartheta\kappa(1-\x)-(\sigma_2+\sigma_3)\beta\kappa(1-\x)\bigg]\rho\right\} \nonumber\\
&+\d_{\x}\d_{\vt}\bigg\{-\zeta(1-2\x)\rho\bigg\} + \d_{\x}\d_{\y}\bigg\{\zeta\vartheta(1-2\x)\rho\bigg\}+ \d_{\x}\d_{\z}\bigg\{-\zeta\vartheta\rho\bigg\} \nonumber\\
&+\d_{\vt}^2\bigg\{\ldots\bigg\} +\d_{\y}^2\bigg\{\ldots\bigg\} + \d_{\z}^2\bigg\{\ldots\bigg\} +\d_{\vt}\d_{\y}\bigg\{\ldots\bigg\} +\d_{\vt}\d_{\z}\bigg\{\ldots\bigg\} + \d_{\y}\d_{\z}\bigg\{\ldots\bigg\},\\
L^{(2)}\rho = \:& \d_{\x}\left\{\left[\frac{\gamma}{2\kappa}\vt\z-\frac{(\sigma_1+\sigma_4)\zeta}{2}(1-\x)\right]\rho\right\}+\d_{\x}^2\bigg\{\frac{\zeta}{4\kappa}\rho\bigg\}+\d_{\vt}\bigg\{\ldots\bigg\}+\d_{\y}\bigg\{\ldots\bigg\}+\d_{\z}\bigg\{\ldots\bigg\}.
\end{align}
%
We employ a time scale separation and focus on a solution evolving on the slow time $\tau = \varepsilon^2t=t/N$ using the Ansatz
\begin{equation*}
\rho(\x,\vt,\y,\z,t) = \rho^{(0)}(\x,\vt,\y,\z,\varepsilon^2t) + \varepsilon\rho^{(1)}(\x,\vt,\y,\z,\varepsilon^2t) + \varepsilon^2\rho^{(2)}(\x,\vt,\y,\z,\varepsilon^2t)+\mathcal{O}(\varepsilon^3).
\end{equation*}
Plugging this into (\ref{eq:FP_re_model2}) and comparing same order terms, we have
\begin{equation}\label{eq:op_model2}
L^{(0)}\rho^{(0)}=0,\quad L^{(0)}\rho^{(1)}=-L^{(1)}\rho^{(0)},\quad L^{(0)}\rho^{(2)}=\d_{\tau}\rho^{(0)}-L^{(1)}\rho^{(1)}-L^{(2)}\rho^{(0)}, 
\end{equation}
which indicates that $\rho^{(0)}$ can be written as
\begin{equation*}
\rho^{(0)}(\x,\vt,\y,\z,\tau) = f(\x,\tau)\hat{\rho}(\vt,\y,\z;\x),
\end{equation*}
with a time independent normal distribution $\hat\rho(\vt,\y,\z;\x)$ in $\vt,\y,\z$ that satisfies $L^{(0)}\hat\rho=0$. 
The function $f(\x,\tau)$ modifies $\hat\rho(\vt,\y,\z;\x)$ 
and represents the time evolution of $\rho^{(0)}$ along the line of stationary points. Integrating the last equation of (\ref{eq:op_model2}) from $-\infty$ to $\infty$ with respect to $\vt,\y,\z$ - the left hand side is a total derivative w.r.t. variables of integration and becomes zero - yields the evolution equation
\begin{equation}\label{eq:evol_model2}
\d_{\tau}f = \int \LL\rho^{(1)}d(\vt,\y,\z) + \int L^{(2)}\r d(\vt,\y,\z).
\end{equation}
We start by computing the second integral. Since all terms that are full derivatives w.r.t. $\vt$, $\y$ or $\z$ vanish upon integration, we have
\begin{equation}
\begin{split}
&\quad\int L^{(2)}\rho^{(0)} d(\vt,\y,\z) \\
&= \d_{\x}\left\{\int\left[\frac{\gamma}{2\kappa}\vt\z+\frac{(-\sigma_1-\sigma_4)\zeta}{2}(1-\x)\right]\rho^{(0)}d(\vt,\y,\z)\right\} + \d_{\x}^2\bigg\{\int \frac{\zeta}{4\kappa}\,\rho^{(0)}d(\vt,\y,\z)\bigg\} \\
&= \d_{\x}\left\{\int\left[\frac{\gamma}{2\kappa}\vt\z\right]\rho^{(0)}d(\vt,\y,\z)\right\} + \d_{\x}\left\{\frac{(-\sigma_1-\sigma_4)\zeta}{2}(1-\x)f\right\} + \d_{\x}^2\left\{\frac{\zeta}{4\kappa}f\right\}.
\end{split}
\end{equation}
We take a closer look at the first integral. 
We aim to write $\vt\z =(L^{(0)})^+h^+-g(\x)$ 
for suitable $h^+$, $g$, where $(L^{(0)})^+$ is the adjoint of $L^{(0)}$. 
If we can do so, then
$$\int \vt\z\rho^{(0)}\, d(\vt,\y,\z)
= \int h^+L^{(0)}\rho^{(0)}\, d(\vt,\y,\z)-g(\x)f(\x,\tau)
=-g(\x)f(\x,\tau)
.$$
To identify $h^+$ and $g$, we reduce the 
problem to linear algebra. 
Define the finite-dimensional vector space (for $k\in\mathbb{N}$)
\begin{equation*}
H_k:=\{P:\mathbb{R}^4\rightarrow\mathbb{R}, P\text{ polynomial homogenous of degree $k$ (w.r.t. variables $\vt,\y,\z$)}\},
\end{equation*}
so that, e.g., $\gamma/2\kappa\;\vt\z\in H_2$, while $\gamma/2\kappa/;\vt\z+c\notin H_2$ for $c\in\mathbb{R}\setminus\{0\}$. Examining $(\L)^+$, we find
\begin{align*}
& (\L)^+h^+ = 0, &\text{for }h^+\in H_0. \\
& (\L)^+h^+ = h\in H_1, &\text{for } h^+\in H_1. \\
& (\L)^+h^+ = h+g\in H_2 \oplus H_0, &\text{for } h^+\in H_2.
\end{align*}
In particular the last observation $(\L)^+H_2\rightarrow H_2 \oplus H_0$,
$h^+\mapsto(h,g)$ 
allows to define an operator 
$M:H_2\rightarrow H_2,$ $h^+\mapsto h$. 
To simplify notation, we identify vectors w.r.t.\ a given basis 
in $H_k$ and the corresponding polynomials.

\begin{prop} \label{MpropModel2}W.r.t.\ the canonical basis 
$(\vt^2,\y^2,\z^2,\vt\y,\vt\z,\y\z)$, the operator $M$ has the 
representation
\begin{equation*}
M = \begin{pmatrix}
-2\gamma\vartheta\kappa & 0 & 0 & \gamma\vartheta^2\kappa & -\gamma\vartheta^2\kappa(1-2\x) & 0 \\
0 & -2(\mu+\zeta) & 0 & \zeta/\vartheta & 0 & 0 \\
0 & 0 & -2(\mu+\zeta) & 0 & 0 & 0 \\
2\zeta/\vartheta& 2\gamma\vartheta^2\kappa & 0 & -(\mu+\zeta+\gamma\vartheta\kappa) & 0 & -\gamma\vartheta^2\kappa(1-2\x) \\
0 & 0 & -2\gamma\vartheta^2\kappa(1-2\x) & 0 & -(\mu+\zeta+\gamma\vartheta\kappa) & \gamma\vartheta^2\kappa \\
0 & 0 & 0 & 0 & \zeta/\vartheta & -2(\mu+\zeta)
\end{pmatrix},
\end{equation*}
and is invertible, if 
all parameters are positive. 
Let $h=(a_1(\x),\ldots,a_6(\x))^T\in H_2$. 
Then, $(L^{(0)})^+(h) = M\,h + g(\x)$, 
where $g(\x)\in H_0$ is uniquely defined by $g(\x)=\hat G(\x)\, h$ and
$\hat G(\x)$ denoting the row-vector
\begin{equation}
\hat G(\x) = (2\zeta\kappa, 2\zeta\vartheta^2\kappa, 2\zeta\vartheta^2\kappa, 
- 2\zeta\vartheta\kappa, 2\zeta\vartheta\kappa(1-2\x), - 2\zeta\vartheta^2\kappa(1-2\x)).
\end{equation} 
For $\tilde h \in H_2$, we find $(L^{(0)})^+(M^{-1}\tilde h) = \tilde h + g(\x)$ 
with $g(\x)=\hat G(\x)\, M^{-1} \tilde h$, so that
$$ \int \tilde h\, \rho(\x,\vt,\y,\z,\tau)\,d(\vt,\y,\z)=-g(\x)\, f(\x,\tau).$$ 
\end{prop}
{\bf Proof: } It is straightforward to obtain the representation of $M$ 
and $g$ by applying $\L$ to the elements of the basis given above. Using, e.g., 
Gau\ss{}-elimination, we find that $M$ is invertible if the parameters 
are positive. In order to obtain the $i'$'th component of $\hat G(\x)$, 
consider the $i$'th entry of the canonical basis $b_i$, compute $(L^{(0)})^+b_i$,
and identify the component that is in $H_0$. E.g., for $i=1$ we find 
$b_1=\vt^2$, and 
$(L^{(0)})^+\vt^2 
= -2\vt\,[(-\zeta/\vartheta)\y+\gamma\vartheta\kappa\vt]+2\zeta\kappa$, 
where $-2\vt\,[(-\zeta/\vartheta)\y+\gamma\vartheta\kappa\vt]\in H_2$ and 
$2\zeta\kappa\in H_0$. 
Hence, $(\hat G(\x))_1=2\zeta\kappa$.
The equation $(L^{(0)})^+(M^{-1}h) = h + g(\x)$ with $g(\x)=\hat G(\x)\, M^{-1} h$  
implies
$$ \int h\, \rho(\x,\vt,\y,\z,\tau)\,d(\vt,\y,\z) 
= 
\int [(L^{(0)})^+(M^{-1}h)- g(\x)]\, \rho(\x,\vt,\y,\z,\tau)\,d(\vt,\y,\z) 
= 0 - g(\x)\, f(\x,\tau).
$$
\qed\par\medskip

Solving the system for $h(\x,\vt,\y,\z)=\gamma/2\kappa\;\vt\z$, we obtain
\begin{equation}
g(\x) = -\frac{\gamma\zeta\vartheta}{2(\mu+\gamma\vartheta)}(1-2\x)
\end{equation}
by using the computer algebra package MAXIMA~\citep{maxima}. In summary, we have
\begin{equation}
\begin{split}
&\quad\int L^{(2)}\rho^{(0)}d(\vt,\y,\z) 
= \d_{\x}\left\{\left(
\frac{\gamma\zeta\vartheta}{2(\mu+\gamma\vartheta)}
+\frac{(-\sigma_1-\sigma_4)\zeta}{2}
\right)
(1-2\x)f(\x,\tau)\right\} + \d_{\x}^2\left\{\frac{\zeta}{4\kappa}f(\x,\tau)\right\}.
\end{split}
\end{equation}
Now, we turn to the computation of the first integral in (\ref{eq:evol_model2}). With
\begin{equation}
h_0^+ := \frac{\zeta}{2\vartheta\kappa(\mu+\zeta)}\left[\z+(1-2\x)\y+\frac{(\beta+\zeta\vartheta)}{\zeta}(1-2\x)\vt\right],
\end{equation}
we obtain
$
(\L)^+h_0^+ = -\zeta/(2\vartheta\kappa)\z-\gamma\vartheta/2(1-2\x)\vt.
$
Remember that full derivatives w.r.t. $\vt$, $\y$ or $\z$ vanish upon integration. So,
\begin{equation}
\begin{split}
&\quad\int \LL\rho^{(1)}d(\vt,\y,\z) \\
&= \d_{\x}\int \left[\frac{-\zeta}{2\vartheta\kappa}\z-\frac{\gamma\vartheta}{2}(1-2\x)\vt\right]\rho^{(1)}d(\vt,\y,\z) = \d_{\x}\int \left[(\L)^+h_0^+\right]\rho^{(1)}d(\vt,\y,\z) \\
&=\d_{\x}\int h_0^+ \left[\L\rho^{(1)}\right]d(\vt,\y,\z) = \d_{\x}\int-h_0^+\left[\LL\rho^{(0)}\right]d(\vt,\y,\z).
\end{split}\label{eqnXX}
\end{equation}
To handle $\int-h_0^+\left[\LL\rho^{(0)}\right]d(\vt,\y,\z)$, 
we use partial integration to move all derivatives in $L^{(1)}$ 
w.r.t.\ $\vt,\y,\z$ from $\rho^{(0)}$ to $h_0^+$ (note that 
$h_0^+$ is linear in these variables, s.t. all second derivatives 
in these variables vanish). Let us consider one of these terms 
occurring in $\int-h_0^+\left[\LL\rho^{(0)}\right]d(\vt,\y,\z)$:
\begin{eqnarray}
\begin{split}
&\int \left[\z+(1-2\x)\y+\frac{(\beta+\zeta\vartheta)}{\zeta}(1-2\x)\vt\right]\,
\partial_{\vt\y}
\bigg\{ \ldots  \bigg\}\, d(\vt,\y,\z) \\
&=
\int \partial_{\vt\y}
\left[\z+(1-2\x)\y+\frac{(\beta+\zeta\vartheta)}{\zeta}(1-2\x)\vt\right]\,
\bigg\{ \ldots  \bigg\}\, d(\vt,\y,\z) 
= 0.
\end{split}\label{eqnXXX}
\end{eqnarray}
 It is not possible to use the same 
procedure for $\x$, as we do not integrate w.r.t.\ $\x$. Here
we use the product rule, e.g.,
\begin{eqnarray*}
\begin{split}
&
\int \left[\z+(1-2\x)\y+\frac{(\beta+\zeta\vartheta)}{\zeta}(1-2\x)\vt\right]\,
 \d_{\x}\left\{\left[\frac{-\zeta}{2\vartheta\kappa}\z-\frac{\gamma\vartheta}{2}(1-2\x)\vt\right]\rho\right\}\, d(\vt,\y,\z) \\
&= 
\d_{x}\,\int \left[\z+(1-2\x)\y+\frac{(\beta+\zeta\vartheta)}{\zeta}(1-2\x)\vt\right]\,
 \left\{\left[\frac{-\zeta}{2\vartheta\kappa}\z-\frac{\gamma\vartheta}{2}(1-2\x)\vt\right]\rho\right\}\, d(\vt,\y,\z) \\
&-
\int \d_{\x}\left[\z+(1-2\x)\y+\frac{(\beta+\zeta\vartheta)}{\zeta}(1-2\x)\vt\right]\,
\left\{\left[\frac{-\zeta}{2\vartheta\kappa}\z-\frac{\gamma\vartheta}{2}(1-2\x)\vt\right]\rho\right\}\, d(\vt,\y,\z). 
\end{split}\label{eqnXXXX}
\end{eqnarray*}
In this way, we obtain 
\begin{equation}
\begin{split}
&\quad\int \LL\rho^{(1)}d(\vt,\y,\z) \\
&\begin{rcases} = \frac{\zeta}{2\vartheta\kappa(\mu+\zeta)}\d_{\x}^2\int \big[ \begin{aligned}[t] &\frac{\gamma\vartheta^2(\mu+\zeta)}{2\zeta}(1-2\x)^2\vt^2+\frac{\zeta}{2\vartheta\kappa}\z^2+\frac{\gamma\vartheta}{2}(1-2\x)^2\vt\y \\
&+\frac{\mu+\gamma\vartheta}{2\kappa}(1-2\x)\vt\z+\frac{\zeta}{2\vartheta\kappa}(1-2\x)\y\z\big]\rho^{(0)}d(\vt,\y,\z) \end{aligned} \end{rcases}=\raisebox{.5pt}{\textcircled{\raisebox{-.9pt} {a}}}\\
&\quad\begin{rcases} +\frac{\zeta}{2\vartheta\kappa(\mu+\zeta)}\d_{\x}\int \left[\gamma\vartheta^2\frac{2\mu+\zeta}{\zeta}(1-2\x)\vt^2+\gamma\vartheta\frac{\mu+\zeta}{\zeta}(1-2\x)\vt\y+\frac{\mu+\zeta-\gamma\vartheta\kappa}{\kappa}\vt\z+\frac{\zeta}{\vartheta\kappa}\y\z\right]\rho^{(0)}d(\vt,\y,\z)\end{rcases}=\raisebox{.5pt}{\textcircled{\raisebox{-.9pt} {b}}}\\
&\quad\begin{rcases} +\d_{\x}\int \left[(\sigma_4+\sigma_1)\frac{\zeta}{2}(1-\x)-(\sigma_4+\sigma_1+\sigma_3+\sigma_2)\frac{\zeta\mu}{\mu+\zeta}\x(1-\x)\right]\rho^{(0)}d(\vt,\y,\z)\\
+\frac{\zeta}{2\vartheta\kappa(\mu+\zeta)}\d_{\x}\int -\beta(1-2\x)\d_{\x}\left\{(1-2\x)\rho^{(0)}\right\}-\d_{\x}\left\{\zeta\vartheta\rho^{(0)}\right\}d(\vt,\y,\z)
\end{rcases}=\raisebox{.5pt}{\textcircled{\raisebox{-.9pt} {c}}}.
\end{split}
\end{equation}
We will treat all three terms separately starting with $\raisebox{.5pt}{\textcircled{\raisebox{-.9pt} {c}}}$.
\begin{equation}
\begin{split}
\raisebox{.5pt}{\textcircled{\raisebox{-.9pt} {c}}} 
&= \d_{\x}\left\{\left[(\sigma_4+\sigma_1)\frac{\zeta}{2}(1-\x)-(\sigma_4+\sigma_1+\sigma_3+\sigma_2)\frac{\zeta\mu}{\mu+\zeta}\x(1-\x)\right]f\right\} \\
&\quad-\frac{\zeta}{2\vartheta\kappa(\mu+\zeta)}\d_{\x}^2\left\{\left[\beta(1-2\x)^2+\zeta\vartheta\right]f\right\}-\frac{\zeta}{2\vartheta\kappa(\mu+\zeta)}\d_{\x}\left\{2\beta(1-2\x)f\right\}.
\end{split}
\end{equation}
To compute $\raisebox{.5pt}{\textcircled{\raisebox{-.9pt} {b}}}$, we proceed as in the computations for $\int L^{(2)}\rho^{(0)}d(\vt,\y,\z)$, that is, we solve
\begin{equation*}
M\begin{pmatrix} a_1(\x) \\ a_2(\x) \\ a_3(\x) \\ a_4(\x) \\ a_5(\x) \\ a_6(\x) \end{pmatrix}
= \underbrace{\begin{pmatrix} \gamma\vartheta^2\frac{2\mu+\zeta}{\zeta}(1-2\x) \\ 0 \\ 0 \\ \gamma\vartheta\frac{\mu+\zeta}{\zeta}(1-2\x) \\ \frac{\mu+\zeta-\gamma\vartheta\kappa}{\kappa} \\ \frac{\zeta}{\vartheta\kappa} \end{pmatrix}}_{\hat{=}h},
\end{equation*}
and find
\begin{equation}
\raisebox{.5pt}{\textcircled{\raisebox{-.9pt} {b}}} = \frac{\zeta}{2\vartheta\kappa(\mu+\zeta)}\d_{\x}\{-g(\x)f(\x,\tau)\} = \frac{\zeta}{2\vartheta\kappa(\mu+\zeta)}\d_{\x}\left\{\left[2\beta-\frac{\gamma\vartheta^2\kappa(\mu+\zeta)}{\mu+\gamma\vartheta}\right](1-2\x)f(\x,\tau)\right\}.
\end{equation}

For $\raisebox{.5pt}{\textcircled{\raisebox{-.9pt} {a}}}$, this can be done similarly, where the system to be solved is

\begin{equation*}
M\begin{pmatrix} a_1(\x) \\ a_2(\x) \\ a_3(\x) \\ a_4(\x) \\ a_5(\x) \\ a_6(\x) \end{pmatrix}
= \underbrace{\begin{pmatrix} \frac{\gamma\vartheta^2(\mu+\zeta)}{2\zeta}(1-2\x)^2 \\ 0 \\ \frac{\zeta}{2\vartheta\kappa} \\ \frac{\gamma\vartheta}{2}(1-2\x)^2 \\ \frac{\mu+\gamma\vartheta}{2\kappa}(1-2\x) \\ \frac{\zeta}{2\vartheta\kappa}(1-2\x) \end{pmatrix}}_{\hat{=}h},
\end{equation*}

and as a result, we have
\begin{equation}
\begin{split}
\raisebox{.5pt}{\textcircled{\raisebox{-.9pt} {a}}} &= \frac{\zeta}{2\vartheta\kappa(\mu+\zeta)}\d_{\x}^2\{-g(\x)f(\x,\tau)\} \\
&= \frac{\zeta}{2\vartheta\kappa(\mu+\zeta)}\d_{\x}^2\left\{\left[\frac{\beta+\zeta\vartheta}{2}(1-2\x)^2-\frac{\zeta^2\vartheta}{2(\mu+\zeta)}\left[(1-2\x)^2-1\right]\right]f(\x,\tau)\right\}.
\end{split}
\end{equation}
With 
$\int \LL\rho^{(1)}d(\vt,\y,\z) = \raisebox{.5pt}{\textcircled{\raisebox{-.9pt} {a}}} +\raisebox{.5pt}{\textcircled{\raisebox{-.9pt} {b}}} + \raisebox{.5pt}{\textcircled{\raisebox{-.9pt} {c}}}$ 
and 
$$ \d_{\tau}f = \int \LL\rho^{(1)}d(\vt,\y,\z)+\int L^{(2)}\rho^{(0)}d(\vt,\y,\z), $$
we obtain
\begin{equation}
\d_{\tau}f = \frac{\zeta}{(1+G)^2}\d_{x}^2\bigg\{\x(1-\x)f\bigg\} + \frac{\zeta(-\sigma_1-\sigma_4-\sigma_2-\sigma_3)}{1+G}\d_{x}\bigg\{\x(1-\x)f\bigg\}.
\end{equation}

\subsection{Fixed population size and fluctuating seed bank}
\label{appendixModel3}
We only present the main steps for this and the following model,
as the computations are lengthy and resemble that of the last model.\par\medskip

Let $p_{i,j,k}(t)=\mathbb{P}(X_t=i,Y_t=j,Z_t=k)$ be the probability distribution of the resulting stochastic process. The corresponding master equation reads:

\begin{equation}
\begin{split}
\dot{p}_{i,j,k}(t) 
= &-\bigg[\zeta N + \frac{\sigma_1}{N}\zeta (N-i) + \beta N - \frac{\sigma_3}{N} \beta (N-i) + \mu (j+k) + \frac{\sigma_2}{N}\mu k\bigg]p_{i,j,k}(t) \\
&+ \bigg[\zeta \frac{i(j+1)}{j+k+1}\bigg]p_{i,j+1,k}(t) 
 + \bigg[\zeta \frac{(i+1)(k+1)}{j+k+1}\bigg]p_{i+1,j,k+1}(t) \\ 
&+ \bigg[(1+\frac{\sigma_1}{N})\zeta\frac{(N-i+1)(j+1)}{j+k+1}\bigg]p_{i-1,j+1,k}(t) \\
&+ \bigg[(1+\frac{\sigma_1}{N})\zeta\frac{(N-i)(k+1)}{j+k+1}\bigg]p_{i,j,k+1}(t) + \bigg[\beta i\bigg]p_{i,j-1,k}(t) + \bigg[\mu (j+1)\bigg]p_{i,j+1,k}(t) \\ 
&+ \bigg[(1-\frac{\sigma_3}{N})\beta (N-i)\bigg]p_{i,j,k-1}(t) + \bigg[(1+\frac{\sigma_2}{N})\mu (k+1)\bigg]p_{i,j,k+1}(t).
\end{split}
\end{equation}

We now obtain the Fokker-Planck equation for the approximating diffusion process as described before. The first step is to transform the system to a quasi-continuous state space by scaling with $N^{-1}$, \textit{i.e.}, defining $x=i/N$, $y=j/N$, $z=k/N$, $h=1/N$ 
and the quasi-continuous density $u(x,y,z,t)=p_{i,j,k}(t)$. The resulting PDE reads

\begin{equation}\label{eq:FP_model1}
\begin{aligned} 
&\d_tu(x,y,z,t) \\=\quad
&  \d_x\left[\left(\zeta x-\zeta\frac{y}{y+z}-\frac{\sigma_1}{N}\zeta\frac{(1-x)y}{y+z}\right)u(x,y,z,t)\right] \\
&+ \d_y\left[\left(\zeta\frac{y}{y+z}+(\mu y -\beta x) +\frac{\sigma_1}{N}\zeta\frac{(1-x)y}{y+z}\right)u(x,y,z,t)\right] \\
&+ \d_z\left[\left(\zeta\frac{z}{y+z}+(\mu z -\beta (1-x)) +\frac{\sigma_1}{N}\zeta\frac{(1-x)z}{y+z} + \frac{\sigma_3}{N}\beta(1-x) +\frac{\sigma_2}{N}\mu z \right)u(x,y,z,t)\right] \\
&+ \frac{1}{2N}\d_x^2\left[\left(\zeta\frac{xz}{y+z}+(1+\frac{\sigma_1}{N})\zeta\frac{(1-x)y}{y+z}\right)u(x,y,z,t)\right]\\
&+ \frac{1}{2N}\d_y^2\left[\left(\zeta\frac{xy}{y+z}+(1+\frac{\sigma_1}{N})\zeta\frac{(1-x)y}{y+z}+\beta x + \mu y\right)u(x,y,z,t)\right]\\
&+ \frac{1}{2N}\d_z^2\left[\left(\zeta\frac{xz}{y+z}+(1+\frac{\sigma_1}{N})\zeta\frac{(1-x)z}{y+z}+(1-\frac{\sigma_3}{N})\beta (1-x) + (1+\frac{\sigma_2}{N})\mu z\right)u(x,y,z,t)\right]\\
&+ \frac{1}{N}\d_x\d_y\left[-(1+\frac{\sigma_1}{N})\zeta\frac{(1-x)y}{y+z}u(x,y,z,t)\right]
+ \frac{1}{N}\d_x\d_z\left[\zeta\frac{xz}{y+z}u(x,y,z,t)\right].
\end{aligned}
\end{equation}

For $N\rightarrow\infty$, we obtain a deterministic
model, governed by the ODEs
\begin{eqnarray}
\begin{aligned}
\dot{x} &= -\zeta x + \zeta\frac{y}{y+z} = -\zeta x\frac{z}{y+z} + \zeta(1-x)\frac{y}{y+z}, \\
\dot{y} &= \beta x - \mu y - \zeta\frac{y}{y+z}, \\
\dot{z} &= \beta(1-x) - \mu z - \zeta\frac{z}{y+z},
\end{aligned}\label{model3Determ}
\end{eqnarray}
which is a neutral competition model since the selection terms vanish in the limit.
\begin{prop}
Let $\vartheta = (\beta-\zeta)/\mu>0$. Then, there is a line of stationary points for (\ref{model3Determ}) in $[0,1]\times\mathbb{R}^2_+$ given by
$$
(x,y,z)=(x,\vartheta x,\vartheta(1-x)), \qquad x\in[0,1].
$$
The line of stationary points is transversally stable. The eigenvectors perpendicular to the line of stationary points (together with the eigenvalues) read
\begin{equation*}
X_1 = \begin{pmatrix}0\\x\\1-x\end{pmatrix},\quad\lambda_1 = -\mu,
\qquad
X_2 = \begin{pmatrix}\zeta\\-\beta\\ \beta\end{pmatrix},\quad\lambda_2 = -\zeta-\beta/\vartheta.
\end{equation*}
\end{prop}
The proof of this proposition is straight forward, along the lines
of the proof of Proposition~\ref{DetLogPopDetSeed}. 
To formulate the inner solution, we introduce local coordinates

\begin{equation}
\x=x, \quad \y = \varepsilon^{-1}(y+z-\vartheta),
\quad \z = \varepsilon^{-1}(y-z-\vartheta(2x-1),\quad \rho(\x,\y,\z,t)=u(x,y,z,t),
\end{equation}

where $\vartheta=(\beta-\zeta)/\mu$ again and $\varepsilon^2=1/N$. Alternatively formulated, we have

\begin{equation*}
x = \x,\quad y = \frac{\varepsilon}{2}(\y+\z)+\vartheta x,
\quad z = \frac{\varepsilon}{2}(\y-\z)+\vartheta(1-x).
\end{equation*}

$\y$ can be thought of as measuring the deviation of the total amount of seeds from its deterministic value $\vartheta$ and $\z$ as measuring the deviation of the allele ratio in seeds from the allele ratio in plants. Both are scaled by $\varepsilon^{-1}$ so we expect them to be of order $\mathcal{O}(1)$. By transforming derivatives, we have
\begin{align*}
\d_x &= \frac{\d\x}{\d x}\d_{\x} + \frac{\d\y}{\d x}\d_{\y} + \frac{\d\z}{\d x}\d_{\z}
= \d_{\x} - 2\varepsilon^{-1}\vartheta\d_{\z},\quad 
\d_y = \varepsilon^{-1}(\d_{\y}+\d_{\z}),\quad
\d_z 
= \varepsilon^{-1}(\d_{\y}-\d_{\z}).
\end{align*}

We now approximate $\d_t\rho(\x,\y,\z,t)$ by transforming all terms on the r.h.s. of (\ref{eq:FP_model1}) and ignoring terms of $\mathcal{O}(\varepsilon^3)$. We find
\begin{equation}\label{eq:FP_res_model1}
\d_t\rho(\x,\y,\z,t)=\left(L^{(0)}+\varepsilon L^{(1)}+\varepsilon^2 L^{(2)}\right)\rho(\x,\y,\z,t)+\mathcal{O}(\varepsilon^3),
\end{equation}

with linear differential operators $L^{(0)}$, $L^{(1)}$ and $L^{(2)}$ that take the form
\begin{align}
L^{(0)}\rho =\hspace*{4pt}&\d_{\y}\bigg\{\mu\y\rho\bigg\}+\d_{\z}\left\{\left[\mu\z+\zeta\left(1+\frac{1}{\vartheta}\right)(\z+\y(1-2\x))\right]\rho\right\}\nonumber \\
&+\d_{\y}^2\bigg\{\beta\rho\bigg\} -\d_{\y}\d_{\z}\bigg\{2\beta(1-2\x)\rho\bigg\} + \d_{\z}^2\left\{\bigg[\beta+4\zeta\vartheta(1+\vartheta)\x(1-\x)\bigg]\rho\right\},\\
L^{(1)}\rho=\hspace*{4pt}
&-\d_{\x}\left\{\frac{\zeta}{2\vartheta}(\z+\y(1-2\x))\rho\right\}
    +\d_{\y}\left\{\bigg[\sigma_1\zeta(1-\x)+\sigma_3\beta(1-\x)+\sigma_2\mu\vartheta(1-\x)\bigg]\rho\right\}\nonumber\\
&-\d_{\z}\left\{\left[\frac{\zeta\y}{\vartheta}\left(1+\frac{1}{\vartheta}\right)(\z+\y(1-2\x)) -\sigma_1\zeta\,(2(\vartheta+1)\x-1)\,(1-\x)+\sigma_3\beta(1-\x)+\sigma_2\mu\vartheta(1-\x)\right]\rho\right\}\nonumber \\
&+\d_{\y}^2\bigg\{\frac{\mu}{2}\y\rho\bigg\} 
 +\d_{\z}^2\left\{\bigg[\zeta(1+\vartheta)(\z+\y(1-2\x))-4\zeta(\vartheta+1)\y\x(1-\x)+\frac{\mu\y}{2}- 2\x\zeta(\z-\y)(1+\vartheta) \bigg]\rho\right\}\nonumber\\
&-\d_{\x}\d_{\z}\bigg\{2(2\vartheta+1)\zeta\x(1-\x)\rho\bigg\} 
 +\d_{\y}\d_{\z}\bigg\{\left(\mu\z+\zeta(1+1/\vartheta)(\z+\y(1-2\x))\right)\rho\bigg\},\\
L^{(2)}\rho =\hspace*{4pt}& \d_{\x}\left\{\left[\frac{\zeta}{2\vartheta^2}\y(\z+\y(1-2\x))
                  -\sigma_1\zeta\x(1-\x)\right]\rho\right\}+\d_{\x}^2\bigg\{\zeta\x(1-\x)\rho\bigg\}+\d_{\y}\bigg\{\ldots\bigg\}+\d_{\z}\bigg\{\ldots\bigg\}.
\end{align}


Before we proceed, we define the functions
\begin{align*}
h_0(\y,\z;\x) \quad=&\quad\frac{-[\z+(1-2\x)\y]}{\mu+\zeta(1+1/\vartheta)},\\
h_1(\y,\z;\x) \quad=&\quad\frac{-[(\z+(1-2\x)\y)\y]}{2\mu+\zeta(1+1/\vartheta)},\\
\begin{split}
h_2(\y,\z;\x) \quad=&\quad\frac{((1-2\x)\y)^2\zeta(1+1/\vartheta)}{2(2\mu+\zeta(1+1/\vartheta))(\mu+\zeta(1+1/\vartheta)}\\ &- \frac{(1-2\x)\y\z\mu}{(2\mu+\zeta(1+1/\vartheta))(\mu+\zeta(1+1/\vartheta)} - \frac{\z^2}{2(\mu+\zeta(1+1/\vartheta))}, \end{split}\\
g(\x) \quad=&\quad4(\x(1-\x))\frac{\beta+\zeta\vartheta(1+\vartheta)}{\mu+\zeta(1+1/\vartheta)},
\end{align*}
and note that 

\begin{align}
(L^{(0)})^+h_0(\y,\z;\x) &= (\z+(1-2\x)\y),\label{relA3e1}\\
(L^{(0)})^+h_1(\y,\z;\x) &= (\z+(1-2\x)\y)\y,\label{relA3e2}\\
(L^{(0)})^+h_2(\y,\z;\x) + g(\x) &= (\z+(1-2\x)\y)\z.\label{relA3e3}
\end{align}
That is, $h_0$ and $h_1$ are eigenfunctions of $(L^{(0)})^+$. 
Observing the system on slow time $\tau=\varepsilon^2t=t/N$, 
we see rapid dynamics for $\tau$ close to zero, $\tau\in\mathcal{O}(1/N)$ 
to be precise. This is the new boundary layer. For $\tau\in\mathcal{O}(1)$, 
only slow drift effects along the equilibrium line should remain. 
Being interested in the outer solution that develops on the slow time scale, 
we make the Ansatz 

\begin{equation*}
\rho(\x,\y,\z,t;\varepsilon) = \rho^{(0)}(\x,\y,\z,\varepsilon^2t) + \varepsilon\rho^{(1)}(\x,\y,\z,\varepsilon^2t) + \varepsilon^2\rho^{(2)}(\x,\y,\z,\varepsilon^2t)+\mathcal{O}(\varepsilon^3),
\end{equation*}

assuming $\d_t\rho=\mathcal{O}(\varepsilon^2)$ to focus on the outer solution and neglect the boundary layer. We plug this Ansatz into (\ref{eq:FP_res_model1}), compare same order terms on both sides and obtain

\begin{equation}\label{eq:Op_model1}
L^{(0)}\rho^{(0)}=0,\quad L^{(0)}\rho^{(1)}=-L^{(1)}\rho^{(0)},\quad L^{(0)}\rho^{(2)}=\d_{\tau}\rho^{(0)}-L^{(1)}\rho^{(1)}-L^{(2)}\rho^{(0)}. 
\end{equation}

We have $\int \y(\z+\y(1-2\x))\rho\,d(\y,\z)
=\int(L^{(0)})^{+}\,h^+_1\rho\,d(\y,\z)=0$, and hence
\begin{eqnarray}
\int L^{(2)}\rho^{(0)}\,d(\y,\z) &=&  
\d_{\x}\int\left[\frac{\zeta}{2\vartheta}\y(\z+\y(1-2\x))-\sigma_1\zeta\x(1-\x)\right]\rho^{(0)}\,d(\y,\z)
+\d_{\x}^2\int\zeta\x(1-\x)\rho^{(0)}\,d(\y,\z)\nonumber\\
&=& 
-\sigma_1\,\zeta\,\d_{\x}\,\bigg(\x(1-\x)\,f(\tau,\x)\bigg)
+ \zeta \d_{\x}^2\bigg(\zeta\x(1-\x)f(\tau,\x)\, \bigg).
\end{eqnarray}
Using the same procedure as in Appendix~\ref{appendixModel2}, 
consider (\ref{eqnXX})-(\ref{eqnXXXX}), to handle 
$\int L^{1}\rho^{1}\,d(\vt,\y,\z)$, and by applying (\ref{relA3e1})~-~(\ref{relA3e3}), we find 
$$
\int L^{1}\rho^{1}\,d(\vt,\y,\z) = T_a+T_b+T_c,
$$
where
\begin{eqnarray}
T_a &=& \d_{\x}\frac{-\zeta^2}{2\vartheta^2((\vartheta+1)\zeta+\mu\vartheta)}\int(\z+(1-2\x)\y)\y\rho^{(0)}\,d(\vt,\y,\z) = 0,\\
T_b &=& \d_{\x}\left(
 \frac{-\zeta \x(1-\x)\,(-\zeta\sigma_1\vartheta - \zeta\sigma_2 + \beta(\sigma_2+\sigma_3))}
      {(\vartheta+1)\zeta+\mu\vartheta}
 \int\rho^{(0)}\,d(\vt,\y,\z)\right),\\
 T_c &=& 
 \d^2_{\x}\left( 
 \frac{\zeta^2}{4\vartheta^2((1+1/\vartheta)\,\zeta+\mu)}
 \int (\z+(1-2\x)\y)\z + (\z+(1-2\x)\y)\,\y\,(1-2\x))\,\rho^{(0)}\,d(\vt,\y,\z)\right)\nonumber\\
 && - \d^2_{\x}\left( 
 \frac{\zeta^2}{\vartheta ((1+1/\vartheta)\,\zeta+\mu)}
 \int (2\vartheta+1)\x(1-\x)\,\rho^{(0)}\,d(\vt,\y,\z)\right)\nonumber\\
 &=&  \d^2_{\x}\left( 
 \frac{-\zeta^2\,[(\vartheta+1)^2\zeta+\mu\vartheta(1+2\vartheta)+\beta)] \x(1-\x) }{(\vartheta((1+1/\vartheta)\zeta+\mu))^2}
 \int \,\rho^{(0)}\,d(\vt,\y,\z)\right).
 \end{eqnarray}
 With
$$ \d_{\tau}f = \int \LL\rho^{(1)}d(\vt,\y,\z)+\int L^{(2)}\rho^{(0)}d(\vt,\y,\z), $$
we find
\begin{eqnarray}
\d_{\tau}f(\x,\tau)  
 &=&   
 \,\frac
 {-\,\zeta[\sigma_1+(1-1/Y)\sigma_2+\sigma_3]}
 {(1+(1-1/Y)G)}
 \d_{\x}\bigg\{\x(1-\x)f(\x,\tau)\bigg\}
 \\
&&+ \frac{\zeta}{(1+(1-1/Y)G)^2}\d_{\x}^2\bigg\{\x(1-\x)
f(\x,\tau)\bigg\}.\nonumber
\end{eqnarray}

\subsection{Logistic population dynamics and fluctuating seed bank}
\label{appendixModel4}

Let $p_{i,j,k,l}(t)=P(X_{1,t}=i,X_{2,t}=j,Y_t=k,Z_t=l)$ be the probability distribution of the resulting stochastic process. The corresponding master equation reads:

\begin{equation}
\begin{split}
\dot{p}_{i,j,k,l}(t) &= -\bigg[ \begin{aligned}[t] &\zeta (i + (1+\frac{\sigma_1}{N})j) + \gamma\frac{N-i-j}{N}(k+(1-\frac{\sigma_4}{N})l) \\
&+ \mu (k + (1+\frac{\sigma_2}{N})l)
 + \beta (i + (1-\frac{\sigma_3}{N})j)\bigg]p_{i,j}(k,l,t) \end{aligned} \\ 
&\quad+ \zeta(i-1)p_{i-1,j,k,l}(t)        
      + (1-\frac{-\sigma_1}{N})\zeta(j-1)p_{i,j-1,k,l}(t) \\
&\quad+ \gamma\,(1-(i+1)/N-j/N)\,(k+1) p_{i-1,j,k+1,l}(t) \\
&\quad+ (1-\frac{\sigma_4}{N})\gamma\,(1-i/N-(j+1)/N)\,(l+1) p_{i,j-1,k,l+1}(t) \\
&\quad+ \beta i\ p_{i,j,k-1,l}(t) + (1-\frac{\sigma_3}{N})\beta j\ p_{i,j,k,l-1}(t)\\
&\quad+ \mu(k+1)p_{i,j,k+1,l}(t) 
      + (1+\frac{\sigma_2}{N}) \mu(l+1)p_{i,j,k,l+1}(t).
\end{split}
\end{equation}

We transform the system to the quasi-continuous state space with rescaled parameters $x_1=i/N$, $x_2=j/N$, $y=k/N$, $z=l/N$, $h=1/N$ and quasi-continuous density $u(x_1,x_2,y,z,t)=p_{i,j,k,l}(t)$. After expanding about $(x_1,x_2,y,z)$ in terms up to order $\mathcal{O}(\varepsilon^2)$, we have a 4-dimensional Fokker-Planck equation again:

\begin{equation}\label{eq:FP_model3}
\begin{aligned} 
\d_tu &=\d_{x_1}\bigg\{\bigg[\zeta x_1-\gamma(1-x_1-x_2)y\bigg]u\bigg\} \\
&\quad+\d_{x_2}\left\{\bigg[(1+\frac{\sigma_1}{N})\zeta x_2-(1-\frac{\sigma_4}{N})\gamma(1-x_1-x_2)z\bigg]u\right\} \\
&\quad+\d_y\left\{\bigg[\mu y -\beta x_1
                                + \gamma(1-x_1-x_2)y\bigg]u\right\} \\
&\quad+\d_z\left\{\bigg[(1+\frac{\sigma_2}{N})\mu z -(1-\frac{\sigma_3}{N})\beta x_2 
                                + (1-\frac{\sigma_4}{N})\gamma(1-x_1-x_2)z\bigg]u\right\} \\
&\quad+\frac{1}{2N}\d_{x_1}^2\left\{\bigg[\zeta x_1+\gamma(1-x_1-x_2)y\bigg]u\right\} \\
&\quad+\frac{1}{2N}\d_{x_2}^2\left\{\bigg[(1+\frac{\sigma_1}{N})\zeta x_2
                                +(1-\frac{\sigma_4}{N})\gamma(1-x_1-x_2)z\bigg]u\right\}\\
&\quad+\frac{1}{2N}\d_{y}^2\left\{\bigg[\mu y+\beta x_1
                                           + \gamma(1-x_1-x_2)y\bigg]u\right\} \\
&\quad+\frac{1}{2N}\d_{z}^2\left\{\bigg[ (1+\frac{\sigma_2}{N})\mu z 
                                + (1-\frac{\sigma_3}{N})\beta x_2
                                + (1-\frac{\sigma_4}{N})\gamma(1-x_1-x_2)z
                                 \bigg]u\right\}\\
&\quad -\frac 1 N \d_{x1}\d_{y}\left\{\bigg[\gamma(1-x_1-x_2)y
                                 \bigg]u\right\}
       -\frac 1 N \d_{x2}\d_{z}\left\{\bigg[(1-\frac{\sigma_4}{N})\gamma(1-x_1-x_2)z
                                 \bigg]u\right\}.\\
\end{aligned}
\end{equation}

The limiting deterministic system for $N\rightarrow\infty$ has dynamics
\begin{eqnarray}
\begin{aligned}
\dot{x_1} &= \gamma(1-x_1-x_2)y -\zeta x_1, \\
\dot{x_2} &= \gamma(1-x_1-x_2)z -\zeta x_2, \\
\dot{y} &= \beta x_1 - \mu y - \gamma(1-x_1-x_2)y,\\
\dot{z} &= \beta x_2 - \mu z - \gamma(1-x_1-x_2)z.
\end{aligned}
\end{eqnarray}
As before, we can show that a line of stable equilibria exists:

\begin{prop}
Let $\vartheta := (\beta-\zeta)/\mu>0$, 
$\kappa := (\gamma\vartheta-\zeta)/\gamma\vartheta \in [0,1]$. 
Then, there is a line of stationary points 
in $[0,\kappa]^2\times\mathbb{R}^2_+$ given by
\begin{align*}
(x_1,x_2,y,z)=(x, \kappa-x, \vartheta x, \vartheta(\kappa-x)), &&x\in[0,\kappa].
\end{align*}

The line of stationary points is transversally stable (locally and globally).
\end{prop}

Since the proof is similar to the proof of Proposition~\ref{DetLogPopDetSeed}, 
it is omitted here.

\subsubsection{Perturbation approximation}
As before, new local variables for the boundary layer around the equilibrium line are defined:
\begin{eqnarray}
x_1&=&\kappa\,\x +\frac 1 2 \varepsilon\,\vt, 
\quad x_2 = \kappa\,(1-\x) +\frac 1 2 \varepsilon\,\vt, \\
\quad y &=& \kappa\,\vartheta\,\x+\frac 1 2 \varepsilon\,(\y+\z+\vartheta\,\vt), 
\quad z =  \kappa\,\vartheta\,(1-\x) +\frac 1 2 \varepsilon\,(\y-\z+\vartheta\,\vt),\nonumber
\end{eqnarray}
where
\begin{eqnarray*}
\x&=&\frac{x_1-x_2+\kappa}{2\kappa}, \quad \vt = \varepsilon^{-1}(x_1+x_2-\kappa), \\
\quad \y &=& \varepsilon^{-1}(y+z-\vartheta(x_1+x_2)), \quad \z = \varepsilon^{-1}(y-z-\vartheta(x_1-x_2)),
\end{eqnarray*}
and $\varepsilon^2=1/N$. 
The considerations for the global behavior are the same as for model 2. We choose the same local variables for the boundary layer around the equilibrium line and transform (\ref{eq:FP_model3}) up to terms of $\mathcal{O}(\varepsilon^3)$ and higher. 
The resulting reparametrized Fokker-Planck equation in the local variables is
\begin{equation}\label{eq:FP_re_model3}
\d_t\rho(\x,\vt,\y,\z,t) = \left(\L+\varepsilon\LL + \varepsilon^2L^{(0)}\right)\rho(\x,\vt,\y,\z,t) + \mathcal{O}(\varepsilon^2),
\end{equation}
with linear differential operators
\begin{align}
L^{(0)}\rho   =\quad& \d_{\vt}\bigg\{\bigg[-\frac{\zeta}{\vartheta}\y 
                    + \gamma\vartheta\kappa\,\vt\bigg]\,\rho\bigg\}
   + \d_{\y} \bigg\{\bigg[(\beta+\zeta\vartheta)\y/\vartheta     
                 - \gamma\vartheta\,(1-\vartheta)\kappa\vt\bigg]\,\rho\bigg\}
                 \nonumber\\
   &+ \d_{\z}  \bigg\{\bigg[ (\beta+\zeta\vartheta)\z/\vartheta     
         + \gamma\vartheta\,(1+\vartheta)\kappa\,(1-2\x)\vt\bigg]\,\rho\bigg\}\nonumber\\
   &+ \d^2_{v} \bigg\{\zeta\,\kappa\,\,\rho\bigg\}
   + \d^2_{\y}\bigg\{\bigg[ \kappa\,(\beta + \zeta\vartheta\,(1+\vartheta))\bigg]\,\rho\bigg\}
   + \d^2_{\z} \bigg\{\bigg[\kappa\,(\beta + \zeta\vartheta\, (1+\vartheta))\bigg]\,\rho\bigg\}\nonumber\\
   &- \d_{\vt}\,\d_{\y} \bigg\{\zeta\,(1+2\vartheta)\kappa \rho\bigg\}
 +\d_{\vt}\,\d_{\z} \bigg\{ \zeta(1+2\vartheta)\kappa\,
                     (1-2\x)\rho\bigg\} \nonumber\\
 &-\d_{\y}\,\d_{\z} \bigg\{ 
    [2\,(1+\vartheta)\vartheta\zeta +2 \beta]\,\kappa\,(1-2\x) 
 \,\rho\bigg\},
\label{model4L0}\\
L^{(1)}\rho=\quad& 
  - \d_{\x} \bigg\{\bigg[
  \frac{\zeta}{2\vartheta\kappa} \,\,\z+\frac{\gamma\vartheta}2 \,\,(1-2\x)\vt 
   \bigg]\rho\bigg\}
 + \d_{\vt} \bigg\{\bigg[
    \vt\gamma (\vartheta\,\vt+\y) + (\sigma_4+\sigma_1)\zeta\kappa(1-\x)
    \bigg]\rho\bigg\}\nonumber\\
 &+\d_{\y} \bigg\{\bigg[ 
    - \vt\gamma(1+\vartheta)(\vartheta\vt+\y) 
     + (\sigma_2+\sigma_3)\beta\kappa(1-\x) 
     - (\sigma_4\,(1+1/\vartheta)+\sigma_1)(1-\x)\zeta\vartheta\kappa
     \bigg]\rho\bigg\}\nonumber\\
 &+\d_{\z} \bigg\{\bigg[ 
       -  \vt\gamma(1+\vartheta)\z 
       - (\sigma_2+\sigma_3)\beta\kappa(1-\x) 
       + (\sigma_4\,(1+1/\vartheta)+\sigma_1)(1-\x)\zeta\vartheta\kappa 
   \bigg]\rho\bigg\}\nonumber\\
 & - \d_{\x}\d_{\vt}\bigg\{ \zeta\,(1-2\x)\rho\bigg\}
  + \d_{\x}\d_{\y}\bigg\{ \zeta(1-2\x)(\vartheta+1/2)\rho\bigg\}
  - \d_{\x}\d_{\z}\bigg\{ \zeta(\vartheta +1/2)\rho\bigg\}\nonumber\\
&+\d_{\vt}^2\bigg\{\ldots\bigg\} +\d_{\y}^2\bigg\{\ldots\bigg\} + \d_{\z}^2\bigg\{\ldots\bigg\} +\d_{\vt}\d_{\y}\bigg\{\ldots\bigg\} +\d_{\vt}\d_{\z}\bigg\{\ldots\bigg\} + \d_{\y}\d_{\z}\bigg\{\ldots\bigg\},     
\label{model4L1}\\
L^{(2)}\rho = \quad& \d_{\x}\left\{\left[\frac{\gamma}{2\kappa}\vt\z-\frac{(\sigma_1+\sigma_4)\zeta}{2}(1-\x)\right]\rho\right\}+\d_{\x}^2\bigg\{\frac{\zeta}{4\kappa}\rho\bigg\}+\d_{\vt}\bigg\{\ldots\bigg\}+\d_{\y}\bigg\{\ldots\bigg\}+\d_{\z}\bigg\{\ldots\bigg\}.
\label{model4L2}
\end{align}

To handle the integrals below, we define an operator 
$M:H_2\rightarrow H_0$ resp.\ $\hat G:H_2\rightarrow H_0$ in  a 
similar way as above (Proposition~\ref{MpropModel2}). 
Recall that we use 
$(\vt^2,\y^2,\z^2,\vt\y,\vt\z,\y\z)$ as the basis in $H_2$.
The operator $M$ has the 
representation
{\footnotesize
\begin{equation*}
M = \begin{pmatrix}
-2\gamma\vartheta\kappa &0                            &0                                    &\gamma\vartheta(1+\vartheta)\kappa                  &-\gamma\vartheta(1+\vartheta)\kappa(1-2\x)          &0\\
0                    &-2(\beta+\zeta\vartheta)/\vartheta &0                                    &\zeta/\vartheta                                  &0                                             &0\\
0                    &0                            &-2(\beta+\zeta\vartheta)/\vartheta         &0                                             &0                                             &0\\
2\zeta/\vartheta        &2\gamma\vartheta(1+\vartheta)\kappa&0                                    &-\gamma\vartheta\kappa-(\beta+\zeta\vartheta)/\vartheta&0                                             &-\gamma\vartheta(1+\vartheta)\kappa(1-2\x)\\
0                    &0                            &-2\gamma\vartheta(1+\vartheta)\kappa(1-2\x)&0                                             &-\gamma\vartheta\kappa-(\beta+\zeta\vartheta)/\vartheta&\gamma\vartheta(1+\vartheta)\kappa\\
0                    &0                            &0                                    &0                                             &\zeta/\vartheta                                  &-2(\beta+\zeta\vartheta)/\vartheta
\end{pmatrix}.
\end{equation*}
}

If we define $g$ via $g(\x)=\hat G(\x)\, h$, we find
\begin{equation*}
\hat G(\x) = \kappa\,\,\,
\bigg(
  2\zeta,\,\,\,
      2(\vartheta(\vartheta+1)\zeta+\beta),\,\,\,
     (2(\vartheta(\vartheta+1)\zeta+\beta),\,\,\,
   - (1+2\vartheta)\zeta,\,\,\,
     (1+2\vartheta)\zeta\,(1-2\x),\,\,\,
   -2(\vartheta(1+\vartheta)\zeta+\beta)(1-2\x)
\bigg).
\end{equation*}

As before, we employ time scale separation and focus on a solution evolving on the slow time $\tau = \varepsilon^2t=t/N$, using the Ansatz

\begin{equation*}
\rho(\x,\vt,\y,\z,t) = \rho^{(0)}(\x,\vt,\y,\z,\varepsilon^2t) + \varepsilon\rho^{(1)}(\x,\vt,\y,\z,\varepsilon^2t) + \varepsilon^2\rho^{(2)}(\x,\vt,\y,\z,\varepsilon^2t)+\mathcal{O}(\varepsilon^2).
\end{equation*}
If we define $h_0^+$ by 
$$ h_0^+ 
=
\frac{\zeta}{2\kappa(\beta+\zeta\vartheta)}
 \bigg[\z+(1-2\x)\y+\frac{\beta+\zeta\vartheta}{\zeta}\,(1-2\x)\, \vt
 \bigg],
$$
we find
$$(L^{(0)})^+h_0^+ =
\frac{\zeta}{2\vartheta\kappa} \,\,\z+\frac{\gamma\vartheta}2 \,\,(1-2\x)\vt.
$$
By now, we have all ingredients together to go along the same route as in Appendix~\ref{appendixModel2}. 
We start with $\int L^{(2)}\rho^{(0)}\, d(\vt,\y,\z)$. 
If we use 
$$
\frac{\gamma}{2\,\kappa}\,\vt\z
=
(L^{(0)})^+
M^{-1}\left(
\frac{\gamma}{2\,\kappa}\,\vt\z\right)
+ 
\frac{\gamma\zeta(\beta+\zeta\vartheta-\beta\vartheta)}
     {2\mu(\beta+\gamma\vartheta^2)}(1-2\x),
$$
we obtain 
\begin{equation}
\int L^{(2)}\rho^{(0)}\, d(\vt,\y,\z) =
\d_{\x}
\left\{\bigg[\frac{\gamma\zeta(\beta+\zeta\vartheta-\beta\vartheta)}
      {2\mu(\beta+\gamma\vartheta^2)}(1-2\x)
      - \frac{(\sigma_1+\sigma_4)\zeta(1-\x)}{2}\bigg]\,f\,\right\}
+ 
\d^2_{\x}\left\{\frac{\zeta}{4\kappa}\,\,f\,\right\}.
\end{equation}
Next, we turn to $\int L^{(1)}\rho^{(1)}\, d(\vt,\y,\z)
=-\int h_0^+\,L^{(1)}\rho^{(0)}\, d(\vt,\y,\z)$
with
\begin{eqnarray*}
h_0^+ 
= 
\frac{\zeta(\z+(1-2\x)\y+(\beta+\zeta\vartheta)(1-2\x)\vt/\zeta)} 
     {(2\kappa(\beta+\zeta\vartheta))}.
\end{eqnarray*}
We integrate by parts w.r.t.\ all derivatives 
$\d_{\vt}$, $\d_{\y}$, $\d_{\z}$, and respectively move the derivatives 
$\d_{\x}$ in front of the integral by means of the chain rule, so that we obtain 
$$-\int h_0^+\,L^{(1)}\rho^{(0)}\, d(\vt,\y,\z)
= \d_{\x}(T_a+T_b+T_c+T_{d,1})+\d^2_{\x}T_{d,2},$$
with
\begin{eqnarray}
T_a &=& \int
\frac{\rho^{(0)}\,\zeta}{2\kappa(\beta+\zeta\vartheta)}\,
\bigg(
         \frac{\gamma\vartheta(\beta+\zeta\vartheta)}{2\zeta}\,\,(1-2\x)^2\vt^2
       + \frac{\zeta\z^2}{2\vartheta\kappa}
       + \frac{\gamma\vartheta}{2}\,\,\,(1-2\x)^2\,\vt\,\y\\
&&\qquad\qquad       + \frac{\beta+\gamma\vartheta^2}{2\vartheta\kappa} \,\,(1-2\x)\vt\z
       + \frac{\zeta}{2\vartheta\kappa}\,(1-2\x)\y\z
       \bigg)
\, d(\vt,\y,\z),\nonumber\\
T_b &=& \int
\frac{\zeta\rho^{(0)}}{2\kappa(\beta+\zeta\vartheta)}\bigg(
   \frac{\gamma\vartheta(2\beta+\zeta\vartheta-\zeta)}{\zeta}\,\,(1-2\x)\vt^2
 + \frac{\gamma\vartheta\,(\mu+\zeta)}{\zeta}\,\,(1-2\x)\vt\y\\
&&\qquad\qquad + \frac{\beta+\zeta\vartheta-\gamma\vartheta(1+\vartheta)\kappa}{\vartheta\kappa}\vt\z
 + \frac{\zeta}{\vartheta\kappa}\,\y\z
\bigg)\, d(\vt,\y,\z),\nonumber\\
T_c &=& \zeta\,\int\bigg(
           \frac 1 2 (\sigma_1+\sigma_4)\,(1-\x)
            -\frac{\beta\,(\sigma_1+\sigma_4(1-\zeta/\beta) 
                 -\sigma_2-\sigma_3)}{\beta+\zeta\vartheta}\,\x\,
                           (1-\x)
           \bigg)\rho^{(0)}
\, d(\vt,\y,\z),\\
T_{d,1} &=& \int\bigg(
\frac{\zeta}{2\kappa(\beta+\zeta\vartheta)}
(\zeta-2\beta)\,(1-2\x)\rho^{(0)}  
\bigg)\, d(\vt,\y,\z),\\
T_{d,2} &=& -\,\int\bigg(
\frac{
-\zeta^2  (2\x(\x-1)-\vartheta)
                  -\zeta\beta  \,\,\,(4\x(1-\x)-1)
                 }{2\kappa(\beta+\zeta\vartheta)}
\bigg)\,\rho^{(0)}\,\, d(\vt,\y,\z).
\end{eqnarray}
In particular, the integrals in $T_a$ and $T_b$ can be transformed 
using the operators $M$ and $\hat G$, 
\begin{eqnarray}
T_a &=& 
\frac{\zeta}{2\kappa(\beta+\zeta\vartheta)}\,\,\,
 \int\bigg(
\frac{\beta+\zeta\vartheta}{2}\,(1-2\x)^2
         -
\frac{2\,\zeta(\beta+\zeta\vartheta(1+\vartheta))}{2(\beta+\zeta\vartheta)}
 \,\,    \x (1-\x)
        )
\bigg)\,\rho^{(0)}\,\, d(\vt,\y,\z),\\
T_b &=& 
\frac{\zeta}{2\kappa(\beta+\zeta\vartheta)}\,\,\,
 \int\bigg(
      2\beta-\zeta
       +\frac{\gamma\kappa\,(\beta+\zeta\vartheta-\beta\vartheta)\,(\beta+\zeta\vartheta)}{\mu(\beta+\gamma\vartheta^2))}
                         \bigg)
                         (1-2\x)
\,\rho^{(0)}\,\, d(\vt,\y,\z).
\end{eqnarray}
Recall that in the present variant of the model 
$\vartheta=(\beta-\zeta)/\mu$, 
$G=\zeta/\mu$ and $Y=\beta/\zeta$.
With
$$ \d_{\tau}f = \int \LL\rho^{(1)}d(\vt,\y,\z)+\int L^{(2)}\rho^{(0)}d(\vt,\y,\z), $$
we find
\begin{equation*}
\d_{\tau}f = \frac{\zeta}{(1+(1-1/Y)G)^2}\d_{x}^2\bigg\{\x(1-\x)f\bigg\} + \frac{\zeta(-\sigma_1-(1-1/Y)\sigma_4-\sigma_2-\sigma_3)}{1+(1-1/Y)G}\d_{x}\bigg\{\x(1-\x)f\bigg\}.
\end{equation*}

\subsection{Fixed population size and deterministic seedbank - Singular perturbation approach}
\label{fixedPopoDetSeedII}

Let $p_{i}(k,l,t)=\mathbb{P}(X_t=i, Y_t\in(k,k+dk), Z_t\in(l,l+dl))$. The master equations 
are given by 
\begin{equation}
\begin{split}
&	\dot p_{i}(k,l,t)	+\nabla\left[ \left(\begin{array}{c}
\beta i-\mu l\\
\beta (1-\sigma_3/N)(N-i)-\mu(1+\sigma_2/N) k
\end{array}\right)p_i(k,l,t)   \right]\\
&= -\left [ 
\zeta i \frac{l}{k+l}+\zeta(1+\sigma_1/N)(N-i)\frac{k}{k+l}
\right] p_{i}(k,l,t)\\
&	+ \left[\zeta\frac{(i+1)l}{k+l}\right] p_{i+1}(k,l,t)
+ \left[\zeta (1+\sigma_1/N)\frac{(N-i+1)k}{k+l}\right] p_{i-1}(k,l,t)
\end{split}
\end{equation}
The usual expansion yields the corresponding Fokker-Planck equation 
($x=i/N$, $y=k/N$, $z=l/N$)
\begin{equation}
\begin{split}
\partial_t u &=
\partial_x\left\{\left[ 
\zeta\frac{xz}{y+z}-\zeta(1+\sigma_1/N)\frac{(1-x)y}{y+z}
\right]u\right\}\\
&	-
\nabla_{y,z}\left\{\left[\begin{array}{c}
\beta x-\mu y\\
\beta (1-\sigma_3/N)(1-x)-\mu(1+\sigma_2/N) z
\end{array}\right]u\right\}\\
& +	\frac 1 {2 N} \partial_x^2\left\{\left[
\zeta\frac{xz}{y+z}+\zeta(1+\sigma_1/N)\frac{(1-x)y}{y+z}
\right]u\right\}
\end{split}
\end{equation}

\subsubsection{Deterministic model}

The drift term of the Fokker-Planck equation 
define the ODE model
\begin{eqnarray*}
	\dot x &=& -\zeta\frac{xz}{x+z}+\zeta\frac{(1-x)y}{y+z}\\
	\dot y &=& \beta x -\mu y\\
	\dot z &=& \beta(1-x)-\mu z
\end{eqnarray*}
with the line of stationary points
$$(x,y,z) = (y,\vartheta x, \vartheta(1-x)),\quad
\mbox{where}\quad \vartheta = \beta/\mu,\quad x\in[0,1].$$

\subsubsection{Dimension reduction by time scale analysis}
We introduce
new coordinates, 
\begin{eqnarray*}
	x = \tilde x, \qquad
	y = \vartheta \x+\frac1 2 \, \eps\,(\y+\z), \quad
	z = \vartheta (1-\x) + \frac 1 2 \, \eps\,(\y-\z),\quad
	\rho(t,\tilde x,\tilde y,\tilde z)=u(t,x,y,z)
\end{eqnarray*}
where, as before, $\vartheta= \beta/\mu$ and $\eps^2=1/N$.
With
\begin{eqnarray}
\x =  x, \qquad
\y = \eps^{-1}(y+z-\vartheta), \quad
\z = \eps^{-1}(y-z+\vartheta (1-2x)),
\end{eqnarray}
we obtain
$$
\partial_x 
= \partial_{\x}-2\eps^{-1}\vartheta\partial_{\z}, \,\,\,
\partial_y 
= \eps^{-1}(\partial_{\y} +\partial_{\z}),\,\,\, 
\partial_z
= \eps^{-1}(\partial_{\y} -\partial_{\z}).
$$
We transform the Fokker-Planck equation, neglecting terms of ${\cal O}(\eps^3)$. 
For $\rho(\x,\y,\z,t;\eps)$ we obtain $\partial_t\rho=L^{(0)}\rho+\eps L^{(1)}\rho+\eps^2L^{(2)}\rho$ with 
\begin{align}
L^{(0)}\rho   = &   \partial_{\y}\bigg[\bigg(\mu\,\y\bigg)\rho\bigg] 
+ \partial_{\z}\bigg[\bigg(\mu\z+\zeta(\z+\y(1-2\x))\bigg)\rho\bigg] 
+ \partial_{\z}^2\bigg[\bigg(4\zeta\vartheta^2\x(1-\x))\bigg)\rho\bigg] 
\end{align}
\begin{align}
L^{(1)}\rho = &
- \partial_{\x}\bigg[\bigg( \zeta(\z+\y(1-2\x))/(2\vartheta)\bigg)\rho\bigg] 
	- \partial_{\y} \bigg[\bigg( - \sigma_3\beta(1-ß\x) - \sigma_2\mu\vartheta(1-\x)\bigg)\rho\bigg]  \nonumber \\
&	- \partial_{\z}\bigg[\bigg(  \zeta(1/\vartheta)\y(\z+\y(1-2\x)) 
- 2\zeta\vartheta\sigma_1\x(1-\x)  ) 
+\beta(\sigma_2+\sigma_3)(1-\x) \bigg)\rho\bigg]\nonumber \\
&	+ \partial_{\z}^2  \bigg[\bigg(  \vartheta\zeta(\y+(1-2\x)\z) -4\zeta\vartheta\x(1-\x)\y\bigg)\rho\bigg] 
	- \partial_{\x}\partial_{\z} \bigg[\bigg( 
4\vartheta\zeta\x(1-\x) \bigg)\rho\bigg] \\
L^{(2)}\rho = &  \partial_{\x} \bigg[\bigg( 
\zeta\y (\z+\y(1-2\x))/(2\vartheta^2) - \sigma_1\zeta\x(1-\x) \bigg)\rho\bigg] 
+  \partial_{\x}^2 \bigg[\bigg( \zeta\x(1-\x) \bigg)\rho\bigg]\nonumber \\
&	+ \partial_{\y}\bigg[\bigg(\cdots\bigg)\rho\bigg]
+ \partial_{\z}\bigg[\bigg(\cdots\bigg)\rho\bigg]
\end{align}
As before, we introduce $\tau = \varepsilon^2t=t/N$, expand $\rho$ w.r.t.\ $\eps$, 
\begin{equation*}
\rho(\x,\y,\z,t) = \rho^{(0)}(\x,\y,\z,\varepsilon^2t) + \varepsilon\rho^{(1)}(\x,\y,\z,\varepsilon^2t) + \varepsilon^2\rho^{(2)}(\x,\y,\z,\varepsilon^2t)+\mathcal{O}(\varepsilon^3)
\end{equation*}
and obtain
\begin{equation}
L^{(0)}\rho^{(0)}=0,\quad L^{(0)}\rho^{(1)}=-L^{(1)}\rho^{(0)},\quad L^{(0)}\rho^{(2)}=\d_{\tau}\rho^{(0)}-L^{(1)}\rho^{(1)}-L^{(2)}\rho^{(0)}.
\end{equation}
The reduced Fokker-Planck equation is given by
\begin{equation}
\d_{\tau}f = \int \LL\rho^{(1)}d(\y,\z) + \int L^{(2)}\r d(\y,\z).
\end{equation}
In the following computations we use that 
\begin{eqnarray}
L^{(0)+}[\y\z+(1-2\x)\y^2))]
&=&
-(2\mu+\zeta)\y \bigg(
\z+\y(1-2\x)
\bigg)
\\
L^{(0)+}(\z+(1-2\x)\y))
&=& -(\mu+\zeta)\,(\z+(1-2\x)\y))\\
L^{(0)+}[(\z+(1-2\x)\y)^2] &=& 
-2(\mu+\zeta)(\z+(1-2\x)\y)^2+8\zeta\vartheta^2\x(1-\x)
\end{eqnarray}
Therewith we find
\begin{eqnarray}
\int L^{(2)}\rho^{(0)} d(\y,zt)
&=&
\frac{\zeta}{2\vartheta^2}\,\partial_{\x}\,\int \y (\z+\y(1-2\x))\,\rho^{(0)}\, d(\y,\z)
- \sigma_1\zeta \partial_{\x} \bigg[\bigg(\x(1-\x) \bigg)f(\x,\tau)\bigg] \nonumber\\
&&	+  \partial_{\x}^2 \bigg[\bigg( \zeta\x(1-\x) \bigg)f(\x,\tau)\bigg]\nonumber\\
&=&
- \sigma_1\zeta \partial_{\x} \bigg[\bigg(\x(1-\x) \bigg)f(\x,\tau)\bigg] 
+  \partial_{\x}^2 \bigg[\bigg( \zeta\x(1-\x) \bigg)f(\x,\tau)\bigg].
\end{eqnarray}
Furthermore, 
\begin{eqnarray}
\int L^{(1)}\rho^{(1)}\, d(\y,\z)
&=&
\frac{-\zeta}{2\vartheta} \partial_{\x}\int \, (\z+\y(1-2\x))\,\rho^{(1)}\, d(\y,\z)\nonumber\\	
&=&
\frac{\zeta}{2\vartheta(\mu+\zeta)} \partial_{\x}\int \, L^{(0)+}(\z+\y(1-2\x))\,\rho^{(1)}\, d(\y,\z)\nonumber	\\
&=&
\frac{-\zeta}{2\vartheta(\mu+\zeta)} \partial_{\x}\int \, (\z+\y(1-2\x))\,L^{(1)}\rho^{(0)}\, d(\y,\z)\nonumber	\\
&=& 
\frac{-\zeta}{2\vartheta(\mu+\zeta)} \partial_{\x}\bigg(
T_1+T_2+T_3+T_4+T_5
\bigg)
\end{eqnarray}
where $T_1$,\ldots,$T_5$ are given by
\begin{eqnarray}
T_1 
&=& \int \, (\z+\y(1-2\x))\,(- \partial_{\x})\bigg[\bigg(
\zeta(\z+\y(1-2\x))/(2\vartheta)
\bigg)\rho^{(0)}\,\bigg]\, d(\y,\z)\nonumber	\\
&=& - \frac{\zeta}{2\vartheta}\bigg[\partial_x \int \,(\z+\y(1-2\x))^2\rho^{(0)}\, d(\y,\z) +2 \int \,\y\,(\z+\y(1-2\x))\rho^{(0)}\, d(\y,\z)\bigg]\nonumber	\\
&=& - \partial_{\x}\frac{\zeta}{2\vartheta}\,\,\,\frac{4\vartheta^2\zeta\x(1-\x)}{\zeta+\mu}\,\, f
= \,\,\frac{-2\vartheta\zeta^2 }{\zeta+\mu}\,\,\partial_{\x}\bigg(\x(1-\x) f\bigg)
\\
T_2 &=& \int \, (\z+\y(1-2\x))\,(- \partial_{\y})\bigg(
- \sigma_3\beta(1-ß\x) - \sigma_2\mu\vartheta(1-\x)
\bigg)\rho^{(0)}\, d(\y,\z)\nonumber	\\
&=& (1-2\x)\bigg(
- (\sigma_2+\sigma_3)\beta(1-ß\x) 
\bigg)\,\,f	
\\
T_3 &=& \int \, (\z+\y(1-2\x))\,(	- \partial_{\z})\bigg(
\zeta(1/\vartheta)\y(\z+\y(1-2\x)) 
- 2\zeta\vartheta\sigma_1\x(1-\x) +\beta(\sigma_2+\sigma_3)(1-\x)
\bigg)\rho^{(0)}\, d(\y,\z)\nonumber	\\
&=& - 2\zeta\vartheta\sigma_1\x(1-\x)\, f +\beta(\sigma_2+\sigma_3)(1-\x)f
\\
T_4 &=& \int \, (\z+\y(1-2\x))\,(\partial_{\z}^2)\bigg(
\vartheta\zeta(\y+(1-2\x)\z) -4\zeta\vartheta\x(1-\x)\y
\bigg)\rho^{(0)}\, d(\y,\z)	= 0
\\
T_5 
&=& \int \,\partial_{\x}\bigg(
4\vartheta\zeta\x(1-\x) 
\bigg)\rho^{(0)}\, d(\y,\z) 
=  \,\partial_{\x}\bigg(
4\vartheta\zeta\x(1-\x) f
\bigg) 
\end{eqnarray}
Adding up the corresponding terms yields the reduced Fokker-Planck equation
\begin{eqnarray}
f_\tau &=& 
\frac{-\zeta\,(\sigma_1+\sigma_2+\sigma_3)}{1+\zeta/\mu}\,\partial_{\x}[\x(1-\x)f]
+ 
\frac{\zeta}{(1+\zeta/\mu)^2}\,\partial_{\x}^2[\x(1-\x)f].
\end{eqnarray}

\end{appendix}

\end{document}